\documentclass[aps,pra,floatfix,superscriptaddress,showpacs,twocolumn]{revtex4}
\usepackage{amsmath}
\usepackage{graphicx}
\def\prn#1{{\left(#1\right)}}

\def\sbrk#1{{\left[#1\right]}}

\newcommand{\abs}[1]{\left|#1\right|}

\def\sbrk#1{{\left[#1\right]}}
\def\abrk#1{{\left\langle#1\right\rangle}}

\def\bra#1{{\left\langle#1\right\vert}}
\def\ket#1{{\left\vert#1\right\rangle}}

\def\prn#1{{\left(#1\right)}}

\def\abs#1{{\left|#1\right|}}

\newcommand{\SSZero}{${\rm 6s^2 \:} ^1{\rm S}_0\:$}

\newcommand{\TDOne}{${\rm 5d6s \:} ^3{\rm D}_1\:$}

\newcommand{\SSZeroToTDOne}{\mbox{${\rm 6s^2}~ ^1{\rm S}_0~\rightarrow ~{\rm 5d6s
}~ ^3{\rm D}_1\:$}}
\newcommand{\TPOneToSSZero}{${\rm 6s6p} ~ ^3{\rm P}_1\rightarrow {\rm
6s^2}~ ^1{\rm S}_0\:$}

\def\TwoByTwo#1#2#3#4{\biggl( \begin{array}{cc} #1 & #2 \\ #3 & #4\end{array}\biggr)}

\begin{document}
\title{Dynamic Stark effect and forbidden-transition spectral lineshapes}
\author{J.\ E.\ Stalnaker}
\altaffiliation[Current Address: ]{National Institute of Standards
and Technology, 325 S. Broadway Boulder, CO 80305-3322}
\affiliation{Department of Physics, University of California,
Berkeley, CA 94720-7300}
\author{D.\ Budker}
\email{budker@socrates.berkeley.edu} \affiliation{Department of
Physics, University of California, Berkeley, CA 94720-7300}
\affiliation{Nuclear Science Division, Lawrence Berkeley National
Laboratory, Berkeley CA 94720}
%\author{D.\ P.\ DeMille}
%\affiliation{Department of Physics, Yale University, New Haven,
%Connecticut 06520}
\author{S.\ J.\ Freedman}
\affiliation{Department of Physics, University of California,
Berkeley, CA 94720-7300} \affiliation{Nuclear Science Division,
Lawrence Berkeley National Laboratory, Berkeley CA 94720}
\author{J.\ S.\ Guzman}
\affiliation{Department of Physics, University of California,
Berkeley, CA 94720-7300}
\author{S.\ M.\ Rochester}
\affiliation{Department of Physics, University of California,
Berkeley, CA 94720-7300}
\author{V.\ V.\ Yashchuk}
%\email{VVYashchuk@lbl.gov}
\affiliation{Advanced Light Source Division, Lawrence Berkeley
National Laboratory, Berkeley CA 94720}

\date{\today}
\begin{abstract}
We report on an experimental and theoretical study of the dynamic
(ac) Stark effect on a forbidden transition. A general framework for
parameterizing and describing off-resonant ac-Stark shifts is
presented. A model is developed to calculate spectral line shapes
resulting from resonant excitation of atoms in an intense standing
light-wave in the presence of off-resonant ac-Stark shifts. The
model is used in the analysis and interpretation of a measurement of
the ac-Stark shifts of the static-electric-field-induced
\SSZeroToTDOne transition at 408 nm in atomic Yb.  The results are
in agreement with estimates of the ac-Stark shift of the transition
under the assumption that the shift is dominated by that of the
\SSZero ground state. A detailed description of the experiment and
analysis is presented. A bi-product of this work is an independent
determination (from the saturation behavior of the 408-nm
transition) of the Stark transition polarizability, which is found
to be in agreement with our earlier measurement. This work is part
of the ongoing effort aimed at a precision measurement of atomic
parity-violation effects in Yb.
\end{abstract}

%32.80.Ys Weak-interaction effects in atoms
%32.70.-n Intensities and shapes of atomic spectral lines
%32.60.+i Zeeman and Stark effects
%32.10.-f Properties of atoms
%42.50.Lc Quantum fluctuations, quantum noise, and quantum jumps
% 33.55.Ad   Optical activity, optical rotation; circular dichroism
% 42.62.Fi   Laser Spectroscopy
% 33.55.Fi   Other magneto-optical and electro-optical effects
%07.55.Ge Magnetometers for magnetic field measurements

\maketitle

\section{Introduction}

Resonant excitation of an atomic transition in an intense
standing-wave light field can lead to surprisingly complex and
non-trivial spectroscopy.  Here we give a detailed experimental and
theoretical analysis of the spectral line shapes resulting from
off-resonant ac-Stark shifts when atoms in an atomic beam are
resonantly excited in such a field.

Spatial variation of the intense standing-wave light field leads to
position- and velocity-dependent ac-Stark shifts. This results in
asymmetric line shapes exhibiting sub-Doppler features.  The
attention to this problem was first drawn by Wieman \emph{et al.\
}\cite{wieman87} who studied the forbidden 6S$\rightarrow$7S
magnetic-dipole (M1) and Stark-induced transition in cesium (Cs) in
the context of their parity-violation experiment (see Ref.
\cite{wood99} and references therein).

In this work, we describe our investigation of this effect on the
\SSZeroToTDOne transition in atomic ytterbium (Yb).  While
conceptually similar to the experiment done in Cs, the atomic
systems and experimental arrangements are sufficiently different so
we are able to study the effects in a different regime.  One of the
main differences in the two works is that the magnitude of the
ac-Stark shifts are significantly larger in Yb than they are in the
Cs transition studied by Wieman \emph{et al.}, while the natural
line width of the forbidden $^1{\rm S}_0\rightarrow^3{\rm D}_1$
transition is an order of magnitude smaller. As a result, we work in
the regime where the ac-Stark shifts often significantly exceed the
natural line width of the transition
\endnote{The work of Ref. \cite{wood97} utilized an optical cavity
with a larger power-build-up factor than that used in Ref.
\cite{wieman87}. Using the power and cavity parameters reported in
Ref. \cite{wood97}, we estimate the maximum ac-Stark shifts were 16
MHz, while the natural line width of the transition is 3.3 MHz.}.
Another key feature of this work is that the Yb transition studied
here is between a $J=0$ ground state and a $J^\prime=1$ excited
state while the Cs transition is $J=1/2\rightarrow J^\prime =1/2$.
Thus, in general, different tensor components of the ac-Stark
polarizability play a role in these two cases. Finally, we are able
to study the effect of the ac-Stark shifts in a regime where
saturation effects become important.

Our interest in this problem is related to the ongoing work in our
laboratory on measuring parity nonconservation (PNC) in atomic Yb
\cite{demille95,bowers99,stalnaker02,budker03}. Detailed
understanding of the line shapes with intense standing-wave
excitation, as well as the magnitude of the ac-Stark shifts of the
transition, are of crucial importance for an accurate measurement of
parity violation in this transition.

In this paper, we develop a theoretical treatment of the line shape
that generalizes and extends the approach of Wieman \emph{et al.}
\cite{wieman87}, and is appropriate for the interpretation of our
experiments with ytterbium.  We also present a measurement of the
ac-Stark shifts for the \SSZeroToTDOne transition and discuss our
theoretical modeling of the line shape.

\section{Theory}

\subsection{Tensor structure of the ac-Stark polarizabilities}
\label{ssec acStarkStructure}

The general tensor structure of ac-Stark polarizabilities has been
discussed by various authors (see e.g.
\cite{happer70,cohentannoudji72,cho97,park01,park02}).  In these
works the ac-Stark shifts are discussed in terms of effective static
electric and magnetic fields.  In this section, we present a
parameterization of the ac-Stark shifts in terms of irreducible
spherical components in a manner analogous to what is commonly done
for dc-Stark shifts.

Ignoring possible time-reversal-invariance-violation effects, there
is no permanent atomic electric-dipole moment and, consequently,
there is no first-order Stark shift. Using time-dependent
second-order perturbation theory and averaging the energy shift over
the period of the light oscillation gives an energy shift for state
$\ket{\gamma, \, J, \, M}$ due to non-resonant light of
\cite{sobelman92}
\begin{align}
\Delta \mathcal{E}\prn{\gamma, J , M}= &{(\varepsilon_0)_i \,
(\varepsilon_0^*)_j \over 4} \sum_{\gamma^\prime, J^\prime,
M^\prime} \bra{\gamma, J, M}d_i \ket{\gamma^\prime, J^\prime,
M^\prime} \times \nonumber \\ & \bra{\gamma^\prime, J^\prime, M^\prime}d_j^*\ket{\gamma, J,  M} \, \times \nonumber \\
& \biggl({1 \over \mathcal{E}\prn{\gamma, J,
M}-\mathcal{E}\prn{\gamma^\prime, J^\prime, M^\prime}+\hbar \,
\omega} \nonumber \\ &
 +{1 \over \mathcal{E}\prn{\gamma, J,
M}-\mathcal{E}\prn{\gamma^\prime, J ^\prime, M^\prime}-\hbar
\omega}\biggr), \label{eq acStarkShiftMatrix}
\end{align}
where the oscillating field (with angular frequency $\omega$) is
given by ${\boldsymbol \varepsilon}(t)={\boldsymbol \varepsilon}_0
\, {\rm cos}(\omega \, t)$, ${\bf d}$ is the dipole operator, and
$\mathcal{E}\prn{\gamma, J , M}$ is the unperturbed energy of state
$\ket{\gamma, \, J, \, M}$. The sum is taken over all of the
opposite-parity atomic energy eigenstates, including the continuum.

Without referring to specific intermediate energy levels, we may say
that the energy shift of a given atomic state, being a scalar
quantity, must result from a contraction of irreducible tensors of
the same rank describing the light and the atom. The light tensors
are bi-linear in the components of the light electric-field
amplitude vector $\vec{\varepsilon}$ and its complex conjugate
$\vec{\varepsilon}^*$. The three irreducible tensor components built
from $\vec{\varepsilon}$ and $\vec{\varepsilon}^*$ are
%--------------------------------------------------------------------------
\begin{align}
&\varepsilon_i \varepsilon_j^* \delta_{ij} && \textrm{scalar}, \label{Eq_light_int}\\
&\!\!\frac{1}{2}\prn{\varepsilon_i \varepsilon_j^*-\varepsilon_j \varepsilon_i^*}&& \textrm{vector}, \label{Eq_Orient}\\
&\frac{1}{2}\prn{\varepsilon_i \varepsilon_j^*+\varepsilon_j
\varepsilon_i^*}-\frac{1}{3}\varepsilon_i \varepsilon_j^*
\delta_{ij} && \textrm{second-rank tensor} \label{Eq_light_align}.
\end{align}
%--------------------------------------------------------------------------
These three tensor components correspond to the intensity,
orientation, and alignment of the light field, respectively. The
vector part of the light tensor, Eq.\ \eqref{Eq_Orient}, can be also
written in terms of the dual ``circular-intensity" vector:
%--------------------------------------------------------------------------
\begin{align}
\frac{1}{2}\prn{\varepsilon_i \varepsilon_j^*-\varepsilon_j
\varepsilon_i^*}=\frac{1}{2}\epsilon_{ijk}V_k,
\end{align}
%--------------------------------------------------------------------------
where
%--------------------------------------------------------------------------
\begin{align}
V_p =
\frac{1}{2}\epsilon_{ijp}\prn{\varepsilon_i\varepsilon_j^*-\varepsilon_j\varepsilon_i^*},\label{Eq_Dual_Circ_vector}
\end{align}
%--------------------------------------------------------------------------
and $\epsilon_{ijk}$ is the totally anti-symmetric tensor. Note that
for the circular intensity vector to be nonzero, the light
polarization vector components cannot all be of the same phase.
Consequently, the circular-intensity vector vanishes in the limit of
a linearly polarized field. As an example of a non-vanishing
circular-intensity vector, for left-circularly polarized light
propagating along $\hat{z}$,
%--------------------------------------------------------------------------
\begin{align}
\vec{\varepsilon}\prn{\sigma^+} =
-\frac{\hat{x}+i\hat{y}}{\sqrt{2}}|\vec{\varepsilon}|,
\end{align}
%--------------------------------------------------------------------------
%--------------------------------------------------------------------------
\begin{align}
\vec{V}\prn{\sigma^+}=-i|\vec{\varepsilon}|^2\hat{z}.
\end{align}
%--------------------------------------------------------------------------

To obtain a shift of an atomic state, three atomic tensor operators
$T^{(\kappa)}$ of ranks $\kappa=0,1,$ and 2, respectively, must be
contracted with the light tensor components given in Eqs.
\eqref{Eq_light_int}-\eqref{Eq_light_align}, so the ac-Stark shift
operator can be written as
%--------------------------------------------------------------------------
\begin{align}
\hat{H}_{ac}= \prn{\varepsilon_i \varepsilon_j^*
\delta_{ij}}T^{(0)}_{ji} + \prn{\varepsilon_i
\varepsilon_j^*-\varepsilon_j
\varepsilon_i^*}T^{(1)}_{ji}\nonumber\\ + \prn{\varepsilon_i
\varepsilon_j^*-\frac{1}{3}\varepsilon_i \varepsilon_j^*
\delta_{ij}}T^{(2)}_{ji}. \label{Eq AC_Stark_Ham}
\end{align}
%--------------------------------------------------------------------------
%In order to find the eigenvalues for this Hamiltonian the tensors
%$T^{(\kappa)}_{i j}$ must be diagonalized with respect to the
%state's magnetic sublevels.
Generally, the eigenstates of the Hamiltonian \eqref{Eq
AC_Stark_Ham} are superpositions of the magnetic sublevels.
However, if mixing of the different magnetic sublevels is
negligible, for example if the magnetic sublevels are split by a
sufficiently strong dc magnetic field, we can find the shifts of the
magnetic sublevels by simply evaluating the diagonal elements of the
above Hamiltonian. In this case, only the tensor operators with the
tensor index $q=0$ contribute to the matrix elements. These contract
with the $q=0$ components of the electric-field tensors. By the
Wigner-Eckart theorem, the atomic matrix elements are proportional
to the matrix elements of operators of the corresponding rank and
index $q$ built from the components of the total angular momentum
$\hat{\vec{J}}$. This results in an ac-Stark shift for the
$\ket{\gamma, \, J,\,M}$ state of
%--------------------------------------------------------------------------
\begin{align}
\Delta \mathcal{E}_{\gamma, J,M} =&\, C_0 |\vec{\varepsilon}|^2 +
C_1 M V_z \nonumber
\\ & + C_2 \prn{M^2- \frac{1}{3}J(J+1)}
\prn{|\varepsilon_z|^2-\frac{1}{3}|\vec{\varepsilon}|^2},
\end{align}
%--------------------------------------------------------------------------
where the constants $C_n$ are determined by the specifics of the
atomic energy levels. Here we used the explicit form of the tensors
of rank up to 2 with $q=0$ (see, for example, Ref. \cite{Var88},
Section 3.2.2) expressed in Cartesian components, for example, that
the $q=0$ component of $T^{(2)}$ is $T^{(2)}_{zz}$.

We choose to define the constants $C_n$ so that the scalar and the
second-rank tensor parts are consistent with the usual definitions
for static electric polarizabilities:
%--------------------------------------------------------------------------
\begin{align}
\Delta \mathcal{E}_{J,M}^{static} =&\, -\frac{\alpha_0}{2}
\vec{E}^{\,2}  \nonumber
\\&-\frac{\alpha_2}{2}\prn{\frac{3M^2-J(J+1)}{J(2J-1)}}
\frac{3E_z^2-\vec{E}^{\,2}}{2}. \label{Eq_DC_Shift_general}
\end{align}
%--------------------------------------------------------------------------
Here $\vec{E}$ is the static electric field. The second-rank tensor
polarizability is defined so that the tensor shift averaged over $M$
is zero and so that, if the electric field is applied along
$\hat{z}$, then $-\alpha_2 E^2/2$ is the tensor shift of the
stretched states $M=\pm J$. We adopt a similar convention for the
vector term in the ac-Stark case: $-\alpha_1|\vec{\varepsilon}|^2/2$
is the vector shift of the $M=J$ state for left circularly polarized
light propagating along $\hat{z}$. Finally, we have a general
expression:
%--------------------------------------------------------------------------
\begin{align}
\Delta \mathcal{E}_{J,M} = &\, -\frac{\alpha_0}{2}
|\vec{\varepsilon}|^{\,2} - i \frac{\alpha_1}{2}\frac{M}{J} V_z
\nonumber
\\&-\frac{\alpha_2}{2}\prn{\frac{3M^2-J(J+1)}{J(2J-1)}}
\frac{3|\varepsilon_z|^2-|\vec{\varepsilon}|^{\,2}}{2}.
\label{Eq_AC_Shift_general}
\end{align}
%--------------------------------------------------------------------------

This analysis allows one to define the scalar, vector, and tensor
ac-Stark polarizabilities, but, we again point out that Eq.\
\eqref{Eq_AC_Shift_general} is only applicable if the mixing of the
magnetic sublevels due to the ac-Stark perturbation is negligible.
If linearly polarized light is applied in the absence of other
fields, the magnetic sublevels will split along an axis of
quantization defined by the polarization of the light with energy
shifts given by Eq.\ \eqref{Eq_AC_Shift_general}.

Equations \eqref{Eq AC_Stark_Ham} and \eqref{Eq_AC_Shift_general}
dictate the selection rules for the scalar, vector, and second-rank
tensor polarizabilities: $\alpha_1$ vanishes for $J=0$ states;
$\alpha_2$ vanishes for $J=0$ and $J=1/2$ states.

\begin{figure}[h]
\includegraphics[width=2.75 in]{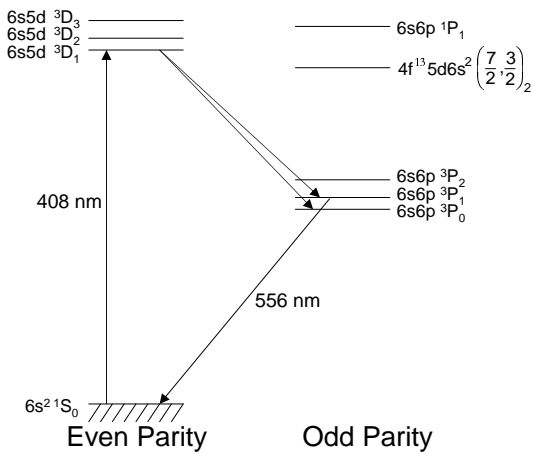}
\caption{Relevant low-lying energy levels and transitions in atomic
Yb. The forbidden transition at 408-nm is driven by a standing-wave
light field in a power-buildup cavity.}\label{fig Energy}
\end{figure}

\subsection{ac-Stark Shifts on a Stark-Induced
Transition} \label{ssec acStarkOnStark}

The transition studied in this work is the highly forbidden
\SSZeroToTDOne transition in atomic Yb (Fig.\ \ref{fig Energy}). In
the absence of external fields this transition occurs through a
highly suppressed magnetic-dipole (M1) amplitude with a value of
\cite{stalnaker02}
\begin{align}
| \abrk{^3{\rm D}_1, M_J | \mu | ^1{\rm S}_0} | = 1.33(8) \times
10^{-4} ~ \mu_0,
\end{align}
where $\mu_0$ is the Bohr magneton.  The above error is reduced from
that quoted in Ref.\ \cite{stalnaker02} due to a reevaluation of the
branching ratios for the decay of the \TDOne state based on the work
of Ref. \cite{porsev99}.

The ongoing PNC experiment relies upon interference of a
parity-violating transition amplitude with a parity-conserving E1
amplitude arising from the application of a dc electric field. Since
the transition of interest is from a $J=0$ state to a $J=1$ state,
the Stark-induced amplitude connecting the ground state and an $M$
sublevel of the \TDOne state can be written as
\begin{align}
A\prn{{\rm E1}_{\rm St}}_M=i \, \beta \prn{{\bf E}\times
\hat{\varepsilon}}_{-M}, \label{eq starkAmp}
\end{align}
where $\prn{{\bf E}\times \hat{\varepsilon}}_{-M}$ is the $-M$
spherical component and $\beta$ is the vector transition
polarizability. The magnitude of $\beta$ was measured to be
\cite{bowers99}
\begin{align}
\abs{\beta} = 2.18(10)\times10^{-8} \: e\,a_0/({\rm V/cm}),
\label{eq beta}
\end{align}
where $e$ is the charge of the electron and $a_0$ is the Bohr
radius.  Again, the error here is reduced from that of Ref.\
\cite{bowers99} in light of Ref. \cite{porsev99}.

For the work discussed here, the Stark-induced E1 amplitude is of
primary interest.  The nature of the Stark-induced transition leads
to a variety of interesting conclusions concerning the experimental
configurations required to access the different ac-Stark-shift
parameters $\alpha_n$. We begin by considering the concrete geometry
employed in this experiment with the $J=0\rightarrow J=1$ transition
under study. In this configuration, the light is linearly polarized
at some angle $\theta$ relative the dc electric field as shown in
Fig.\ \ref{fig_Exp_Arr}.  We define our coordinate system so that
the axis of quantization is along the electric field and the
polarization is nominally in the $x$-$z$ plane (effects of
misalignments are discussed in Section \ref{sec acStarkAnalysis}).
%\begin{figure}
%\includegraphics[width=2.5 in]{PBCGeometry.eps}
%\caption{Orientation of fields for ac-Stark experiment. The axis of
%the power-buildup cavity is aligned along \textbf{y}. The nominal
%direction of the atomic beam (\textbf{v}) is along \textbf{x}.}
%\label{fig acStarkGeom}
%\end{figure}

From Eq.\ \eqref{eq starkAmp} we see that with this geometry, the
light excites the state
\begin{align}
\ket{^3{\rm D}_1, M_y=0}={1\over \sqrt{2}}\prn{\ket{^3{\rm D}_1,
M_z=+1}+\ket{^3{\rm D}_1, M_z=-1}},\label{Eq_M_y=0}
\end{align}
which is a state aligned  along the $y$ axis, regardless of the
polarization angle $\theta$.

%It is possible to show
%% (for example, by explicitly considering the
%% effect of opposite-parity states with $J=0,1,2$)
%that this state is an energy eigenstate of both the dc-Stark
%perturbation and the ac-Stark perturbation, regardless of the angle
%of polarization.
This state is an eigenstate of both the dc- and ac-Stark
perturbations, as can be seen by considering their Hamiltonians,
which are of the form of Eq. \eqref{Eq AC_Stark_Ham}. Because the
electric field vector E does not have a $y$ component, the
Hamiltonian \eqref{Eq AC_Stark_Ham} is a linear combination of the
operator products that transform as $J_iJ_j$ with $i,j=x,z$. Taking
the quantization axis along $y$, the matrix elements of these
products can be written in terms of the matrix elements of products
of pairs of the raising and lowering operators. Such products will
only mix states with $\Delta M_y=0,2$, so the $M_y=0$ state is not
mixed with $M_y=1,-1$.
The $M_y=0$ state remains an energy eigenstate of both interactions
acting simultaneously and its energy does not depend on the angle of
the polarization of the light relative to the dc electric field.
%Diagonalizing the Hamiltonian resulting from the
%interaction of the dc- and ac-Stark shifts gives
Based on the Eqs. \eqref{Eq_M_y=0}, \eqref{Eq_DC_Shift_general}, and
\eqref{Eq_AC_Shift_general}, the energy shift of this state is
\begin{align}
\Delta \mathcal{E}\prn{^3{\rm D}_1, M_y=0} = -&\prn{{1\over 2}\,
\alpha_0^{ac} +{1 \over 2}\, \alpha_2^{ac}}\varepsilon^2 \nonumber
\\ -&\prn{{1\over 2}\, \alpha_0^{dc} + {1 \over 2}\,
\alpha_2^{dc}}E^2. \label{eq eMY}
\end{align}
Here we have used the superscripts $ac$ and $dc$ to distinguish
between the static and dynamic polarizabilities (we omit the
superscripts where the meaning of a symbol is clear from the
context). We conclude that with the current configuration of fields,
the experiment is only sensitive to the combination of ac-Stark
parameters defined in Eq.\ \eqref{eq eMY}. However, it is possible
to observe different components of the ac-Stark shifts using
different configurations of fields. While not directly relevant for
the measurement described here, we briefly discuss the modifications
which would be necessary to isolate the scalar, vector, and tensor
polarizabilities.

Perhaps the simplest modification one can envision in order to
separate the scalar and tensor shifts is to apply a strong magnetic
field along the $z$ axis.  This provides the strong leading field
needed to make Eq.\ \eqref{Eq_AC_Shift_general} valid.  In this
case, the $M_z=-1$ and $M_z=+1$ magnetic sublevels are excited
independently.  As can be seen from Eq.\
\eqref{Eq_AC_Shift_general}, the ac-Stark shift of the two states
will depend on the angle of the polarization of the light with
respect to the dc-electric field. Measuring the shift at different
polarization angles then allows one to separate the scalar and
tensor components.

In order to measure the vector ac-Stark shift parameter, one must
use circularly polarized light since the shift depends on ${\rm
Im}\prn{\varepsilon_x^* \varepsilon_y}$ [see Eqs.
\eqref{Eq_Dual_Circ_vector} and \eqref{Eq_AC_Shift_general}].
Furthermore, the light must propagate non-orthogonally to the axis
of quantization in order to have a nonzero component of the
polarization-intensity vector along the quantization axis. These
conditions can be achieved by
%applying a large magnetic field along
%the $y$ axis of the arrangement shown in Fig.\ \ref{fig acStarkGeom}
%and looking for a difference in the ac-Stark shift for the $M_y=+1$
%and $M_y=-1$ components. \textbf{Dima says: I am not sure I
%understand this. If by the geometry of Fig.\ \ref{fig acStarkGeom}
%we mean the same as on the figure, except $\varepsilon$ circularly
%polarized, and a B-field applied along y, then isn't it true that we
%will still be exciting just the $M_y=0$ sublevel ?} Alternatively,
%one could
using circularly polarized light and applying the electric field
along the axis of light propagation. With such an arrangement, only
one magnetic sublevel, either $M_y = +1 $ or $M_y = -1$, is excited
and the shift of the level would be a combination of the scalar,
vector, and tensor shifts.

\subsection{Estimate of Ground-State Shift}
The shift of the transition frequency is due to shifts in both the
${\rm 6s^2 }\: ^1{\rm S}_0$ ground state and ${\rm 5d6s} \: ^3{\rm
D}_1$ excited state. Since the ground state has angular momentum
$J=0$, it can only have a scalar shift. In contrast, the excited
state can have all scalar, vector, and tensor components.

The ac-Stark shift is resonantly enhanced for eigenstates with
energy separations nearby the frequency of the light.  In addition,
the effect depends quadratically on the dipole coupling of the
states involved. Therefore, one can expect, for light resonant with
the \SSZeroToTDOne transition, the shift of the \SSZero state to be
dominated by the ${\rm 6s6p} \: ^1{\rm P}_1$ state, since it is the
nearest opposite-parity state to the \TDOne state and has a strong
dipole coupling to the ground state.

The shift of the ground state can be estimated using Eq.\ \eqref{eq
acStarkShiftMatrix} and reducing the sum to include only the
$\ket{{\rm 5d6s} \: ^1{\rm P}_1}$ state.  Using the energy
separation of the $^1{\rm P}_1$ and $^3{\rm D}_1$ states, $579 \:
{\rm cm^{-1}}$, and the matrix element $\abs{\bra{{\rm 6s6p} \;
^1{\rm P}_1,{\rm M_z}=0}d_z\ket{{\rm 6s^2} \; ^1{\rm S}_0}}=2.4(1)
\:e \, a_0$, as determined from the lifetime of the ${\rm 6s6p} \:
^1{\rm P}_1$ state \cite{blagoev94}, we find
\begin{align}
\alpha_0\prn{408 \, {\rm nm}, \: ^1{\rm S}_0} \approx 0.28 \; {\rm
{Hz\over\prn{V/cm}^2}}.\label{Eq_alpha_gs_estimate}
\end{align}

The shift of the \TDOne state is significantly harder to estimate.
It is possible that there is an odd-parity eigenstate with energy
close to twice the energy of a 408-nm photon, which can lead to
resonantly enhanced shifts of the \TDOne state. The energy spectrum
in this region (which is below the Yb ionization limit) is very
dense due to the excitation of 4f orbitals \cite{martin78}. While
the ac-Stark shifts of the \TDOne state due to the known energy
eigenstates are estimated to lead to shifts which are significantly
smaller than that of the ground state, the knowledge of the energy
spectrum is far from complete in this region. This provides one of
the motivations for determining the ac-Stark polarizabilities
experimentally.

\section{Experimental Apparatus and Procedures}
\label{sec acStarkExpApp}

\begin{figure}
\centerline{\includegraphics[width=3. in]{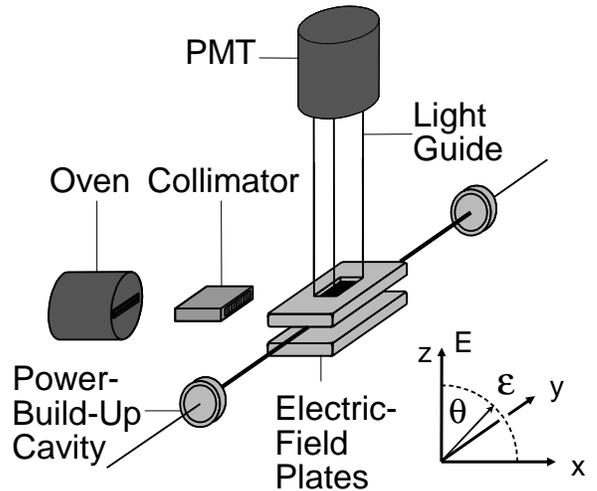}}
\caption{A schematic of the arrangement and orientation of fields
for the ac-Stark experiment.} \label{fig_Exp_Arr}
\end{figure}

A schematic of the experimental arrangement in presented in Fig.
\ref{fig_Exp_Arr}. Atoms were resonantly excited on the
\SSZeroToTDOne Stark-induced transition inside an optical
power-build-up cavity. Fluorescence at 556 nm from the subsequent
\TPOneToSSZero transition (see Fig. \ref{fig Energy}) was recorded
as the laser frequency was scanned over the resonance. The power and
polarization of the excitation light were varied.  A numerical
calculation based on the density-matrix formalism was used to
generate spectral line shapes for different values of the ac-Stark
polarizabilities.  The data were fit to these line shapes in order
to extract a combination of the ac-Stark polarizabilities for the
transition.

\subsection{Vacuum and Atomic Source}

A stainless-steel oven with a multi-channel nozzle was used to
create an effusive beam of Yb atoms inside of a vacuum chamber with
a residual background pressure of $\approx 5 \times 10^{-6}~{\rm
Torr}$. The oven was heated with coiled tantalum-wire heaters placed
inside ceramic tubes (alumina AD-998; Coors Ceramic). The oven was
operated with the front $\approx 100~^{\circ}{\rm C}$ hotter than
the rear to avoid clogging of the nozzle.  The typical operating
temperature was $ 500~^{\circ}{\rm C}$ in the rear of the oven,
corresponding to a atomic beam density of $2.5 \times 10^{14} ~ {\rm
atoms / cm^3}$ inside the oven. A downstream external-vane
collimator was used to reduce the Doppler width to $\Gamma_D \approx
12 \, {\rm MHz}$.  The transparency of the collimator is estimated
to be $\approx 90 \%$ in the forward direction. The collimator was
mounted on a movable platform, allowing precise alignment of the
angle of the collimator relative to the atomic beam during the
experiment. The collimator was heated using tantalum wire heaters to
$\approx 350^\circ {\rm C}$ to prevent clogging.

\subsection{Electric Field} \label{ssec efield}

The dc electric field was produced by applying voltage to a pair of
plates which were the same as those used in Refs.\
\cite{bowers99,stalnaker02}. The plates were made of stainless steel
and were $0.79$~cm thick, $8.9$~cm along the direction of the laser
beam, and $3.2$~cm in the direction of the atomic beam.  The edges
of the plates were rounded to a $0.3$~cm radius and electropolished
to avoid discharge. The plates were oriented so as to provide a
vertical electric field and were separated with two Delrin spacers.
The separation of the plates was $1.016$~cm.  A depression was cut
into the top electrode and an array of holes was drilled into the
plate in order to collect the fluorescence (see Refs.\
\cite{bowers99,stalnaker02} for further details).

For the work described in Ref.\ \cite{bowers99}, the effect of the
holes on the magnitude of the electric field was calculated. It was
found that the holes reduced the electric field by less than $1\%$.
The exact correction was not determined due to uncertainties in the
surface charges of the dielectric light guide that was placed into
the depression to measure the fluorescence.  We therefore,
conservatively estimate a $1\%$ error in our knowledge of the
magnitude of the electric field.  The value of the applied field was
monitored with a precision high-voltage divider.

\subsection{Residual Magnetic Field} \label{ssec bfield}

The magnetic field from the earth was partially compensated using
three sets of magnetic coils wrapped around the vacuum chamber.  We
estimate that the residual magnetic field was trimmed down to a
level of $\approx 50~{\rm mG}$.

\subsection{Power-Build Up Cavity}

The cavity was designed as a symmetric two-mirror resonator with
mirror radii of curvature of 25~cm. The cavity-mirror separation was
26.01(6)~cm and the light was coupled in to the ${\rm TEM}_{00}$
transverse cavity mode.

The cavity consisted of two high-quality mirrors mounted on
precision optical mounts.  The mirror mounts were attached to a
Super-Invar rod supported by lead blocks designed to damp
vibrations.
%
%For the present experiment, it is important to know the intensity of
%the light in the cavity. Therefore, we have carried out an
%experimental characterization of the cavity.

The transmission of the mirrors was measured directly using a
Newport 841-PE power meter with an 818-SL head.  The incident power
was measured with a calibrated attenuator (833-SL); the attenuator
was removed for the measurement of the power transmitted. The power
meter is calibrated to $\pm 2\%$ both with and without the
attenuator. The transmission of the mirrors was measured to be
\begin{align}
T=5.08(7)_{Stat}(14)_{Calib} \times 10^{-4}, \label{eq mirrorTrans}
\end{align}
where the first error is from the variation of the transmission
measurements and the second is the error from the calibration.  The
transmission was measured for both mirrors and found to be the same.

The finesse of the cavity was measured using the cavity-ring-down
method \cite{anderson84}.  Using the experimentally measured
ring-down time
\begin{align}
\tau = 1.17(2) \: {\rm \mu s},
\end{align}
we have a cavity finesse of
\begin{align}
\mathcal{F} \approx {2 \, \pi \, c \over 2\, L} \, \tau = 4240(70),
\end{align}
where $c$ is the speed of light and $L=26.01(6)\:{\rm cm}$ is the
length of the cavity.

Using the measured finesse, we calculate a value for the
reflectivity of the mirrors of
\begin{align}
1-R= 7.41(12) \times 10^{-4}.
\end{align}
The total mirror loss, due to absorption and scattering, per bounce
is estimated to be
\begin{align}
l=1-R-T =2.4(2) \times 10^{-4}.
\end{align}

The amount of loss limits the fraction of power which can be
transmitted through the cavity.  The fraction of the power coupled
into the cavity that is transmitted is given by \cite{hood01}
\begin{align}
{P_{T} \over P_C} & = T^2 \, \prn{\mathcal{F} \over \pi}^2 ={T^2 \over \prn{l+T}^2} \\
&=0.46(13).
\end{align}

In order to determine the amount of power coupled into the cavity we
monitored the light coming off of the input coupler of the cavity.
This light consists of light that is not mode matched to the cavity
as well as light due to an imperfect cancelation of the cavity
leakage field with the light rejected from the cavity mirror.  If we
couple a fraction of light $\epsilon$ into the cavity, $P_C=\epsilon
P_{\rm in}$, then the power coming back from the input coupler is
\cite{hood01}
\begin{align}
P_R= \epsilon \, P_{\rm in} \, l^2 \, \prn{\mathcal{F} \over
\pi}^2+(1-\epsilon)\, P_{\rm in}.
\end{align}
The first term in this equation is due to the imperfect cancelation
of the leakage field with the light directly reflected off of the
mirror and represents the fundamental limitation on the amount of
light that can be coupled into the cavity.  The second term is the
light that is rejected due to imperfect mode matching.

The amount of rejected light when the cavity was on resonance was
$\approx 60\%$ of that off resonance.  Using this value and the
above determinations of $l$ and $\mathcal{F}$ we find
\begin{align}
\epsilon = 0.44.
\end{align}
We believe this coupling efficiency was limited by our inability to
completely compensate the astigmatism of the laser beam.

Combining the above results we find that the fraction of incident
power that is transmitted through the cavity is
\begin{align}
{P_T \over P_{\rm in}} \approx 0.18.
\end{align}
This is in reasonable agreement with the experimentally measured
0.16.

During the experiment, the power transmitted through the
power-build-up cavity was recorded with a photodiode and the
photodiode voltage was calibrated with the same calibrated power
meter used to measure the transmission of the mirror.

\subsection{Laser-Frequency Locking and Stabilization}
\label{ssec acLaserLock}

Figure \ref{fig lockingDiagram} shows a block diagram of the laser
locking system.  A Coherent $899$ Ti:Saphire laser was pumped by
$\approx~\!\!12~{\rm W}$ from a Spectra Physics 2080 argon-ion laser
operating on all lines. The Ti:Sapphire laser produced
$\approx~\!\!1.2~{\rm W}$ of light at $816~{\rm nm}$.  This light
was frequency doubled using a commercial bow-tie resonator with a
Lithium-Triborate crystal (Laser Analytical Systems Wavetrain cw) to
produce the 408-nm light needed to excite the transition. The output
of the frequency doubler was $\approx 80 ~{\rm mW}$.

\begin{figure}
\centerline{\includegraphics[width=3. in]{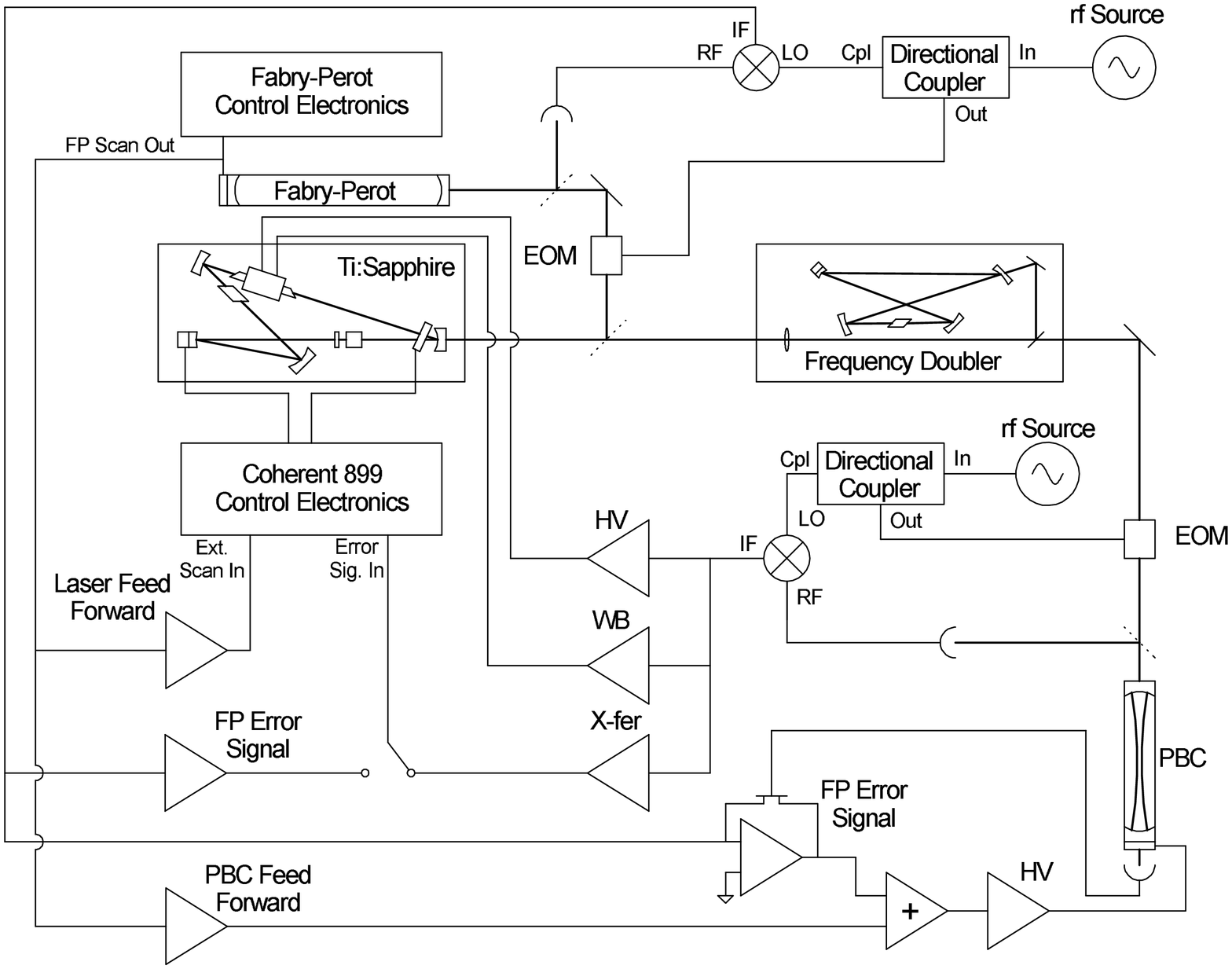}}
\caption{Block diagram of laser frequency stabilization.} \label{fig
lockingDiagram}
\end{figure}

In order to increase the bandwidth of the frequency lock and narrow
the line width of the laser to match that of the power-build-up
cavity, an EOM was placed inside the laser cavity. The EOM used was
a double-crystal assembly from LINOS Optics (PM 25 IR). The EOM was
driven by a homemade electronics circuit.

The laser was locked to the power-build-up cavity using the
fm-sideband technique \cite{drever83}. The laser beam passed through
an EOM (ConOptics 370) to which rf power of 4~W at 28 MHz was
coupled via a resonant LC circuit. The laser beam was coupled on
axis to the power-build-up cavity. The beam passed through a glass
plate which picked off a small fraction of the light rejected from
the input coupler of the cavity. This light was sent to an amplified
fast photodiode (New Focus 1801, bandwidth 125~MHz). The signal from
the photodiode was amplified and mixed down with the local
oscillator to produce the error signal used to control the laser
frequency.

In order to improve absolute frequency stability, the resonant
frequency of the cavity was locked to a hermetically sealed,
temperature-stabilized confocal Fabry-Perot interferometer at 816~nm
(Burleigh CFT-500) with a free-spectral range of 150~MHz using a
similar fm-sideband lock as that described above. The error signal
from this locking system was sent to a high-voltage amplifier which
controlled the piezo-ceramic-mounted mirror of the power-build-up
cavity. Thus, the stable cavity provided the master frequency, with
the power-build-up cavity serving as the secondary master for the
laser. The frequency of the laser system was scanned over the
profile of the atomic resonance by scanning the primary master
Fabry-Perot cavity.  Feed forward was sent to both the
power-build-up cavity and the laser to help them track the frequency
scans.
%\textbf{Once locked, the system typically maintained the lock
%for over an hour before minor adjustments were necessary.}

\subsection{Fluorescence Detection} \label{sssec fluorDet}

The number of atoms excited to the \TDOne state was monitored by
observing the fluorescence of the ${\rm 6s6p} \: ^3{\rm P}_1
\rightarrow {\rm 6s^2} \: ^1{\rm S}_0$ transition at 556~nm. A
lucite light guide was used to send the fluorescent light to a
photomultiplier tube (PMT, Burle 8850) outside of the vacuum
chamber. A narrow-band interference filter, centered around 560 nm,
with a pass band of 10 nm was placed on the front of the
photomultiplier tube to reduce the scattered light reaching the PMT.
The overall detection efficiency was such that $\sim 0.05\%$ of the
atoms undergoing the 408-nm transition produced photoelectrons at
the PMT photocathode. The photocathode was operated at $-1750$ V.
The gain of the tube was measured to be $\approx 2\times 10^6$ at
this voltage. The PMT current was fed into a Stanford Research
Systems current preamplifier (SRS 570). The voltage output was
recorded with a digital oscilloscope and saved to a computer.

\subsection{Frequency Reference}

%The relative frequency of the laser light was determined using a
%homemade Fabry-Perot interferometer operating at 408~nm.
Since the experiment relies on a detailed understanding of the
spectral line shape of the transition over a region of $\approx
100$~MHz, a frequency reference with closely spaced frequency
markers is needed. To this end, we have constructed a Fabry-Perot
interferometer operating at 408~nm. The mirror spacing was chosen so
that the transverse cavity modes overlap at frequency intervals of
\begin{align}
\Delta\nu_{res}={c \over 2\, N \, L},
\end{align}
with $N=7$ \cite{budker00}.  The interferometer mirrors each have a
radius of curvature of $R~\!\!\!=~\!\!\!50 \,{\rm cm}$, and are
separated by 38.9~cm.  This results in a spacing between the cavity
resonances of 55.12 MHz.

\section{Calculation of Line Shape} \label{sect calc}
\subsection{Numerical Modeling}

For a $J=0 \rightarrow J^\prime=1$ transition, there are generally
four magnetic sublevels which must be included in a calculation.
However, as discussed in Section \ref{ssec acStarkOnStark}, the
orientations of fields used in this experiment lead to excitation of
a single energy eigenstate which is a superposition of the $M_y=\pm
1$ sublevels. Thus, we can neglect the structure of the upper state
and treat the excitation of the atoms as a two-level problem.

The calculation was done using the density-matrix formalism. The
Hamiltonian consists of the usual dipole coupling between the two
states with an additional term along the diagonal which describes
the energy shifts due to the ac-Stark effect
\begin{align}
H = \TwoByTwo{0}{d \, \varepsilon(\mathbf{x}, \, t)}{d \,
\varepsilon(\mathbf{x}, \, t)}{\omega_0-{1\over 2}\, \alpha \,
\varepsilon^2(\mathbf{x})}.
\end{align}
Here, $d$ is the dipole matrix element coupling the states, $\alpha$
is the ac-Stark parameter, and $\varepsilon(\mathbf{x},t) =
\varepsilon(\mathbf{x}) {\rm cos}(\omega t)$ is the oscillating
electric field of the standing wave.  We neglect the small residual
running wave component of the field inside the cavity as it is
$\approx 1000$ times smaller than the standing wave component.  The
function $\varepsilon({\bf x})$ is the amplitude of the oscillating
electric field as a function of position. Since we are considering a
two-level system, the tensor structure of the ac-Stark interaction
is not included.  In keeping with the discussion of Section
\ref{ssec acStarkOnStark}, the $\alpha$ which appears here is given
by
\begin{align}
&\alpha=\\ &\alpha_0(408 \, {\rm nm}, \: ^3{\rm D}_1)+ \,
\alpha_2(408 \, {\rm nm}, \: ^3{\rm D}_1)-\alpha_0(408 \, {\rm nm},
\: ^1{\rm S}_0).\nonumber
\end{align}

The spatial variation of the electric field translates into a
temporal variation as an atom passes through the laser beam. The
electric-field amplitude inside the cavity is a standing wave with a
fundamental-Gaussian profile
\begin{align}
\varepsilon(\mathbf{x}) =\varepsilon_0 \, e^{-{(x^2+z^2)\over
r_0^2}} {\rm cos}(2 \pi \, {y\over \lambda}),
\end{align}
where $r_0$ is the electric-field radius of the Gaussian (following
the convention of Ref. \cite{siegman86}), and $\lambda$ is the
wavelength of the light. The nominal direction of the atomic
velocity is along $\hat x$. An atom with velocity
\begin{align}
\mathbf{v} = v_x \, \hat x + v_y \, \hat y + v_z \, \hat z,
\end{align}
is subjected to an electric-field amplitude of
\begin{align}
\varepsilon(t)&=\varepsilon_0 \, e^{-{(v_x \, t)^2 +(z_0+v_z \,
t)^2\over
r_0^2}}{\rm cos}\sbrk{{2 \pi \over \lambda} (y_0+v_y \, t)} \nonumber \\
&\approx \varepsilon_0 \, e^{-{(v_x\, t)^2 + z_0^2\over r_0^2}}{\rm
cos}\sbrk{{2 \pi \over \lambda} (y_0+v_y \, t)}, \label{eq epsilont}
\end{align}
where we have chosen $t=0$ to be when the atom is in the middle of
the laser beam.  We have neglected $v_z\, t$ since the change in $z$
as an atom crosses the laser beam is small. The largest angle an
atom can have in the vertical direction and still intersect the
laser beam is
\begin{align}
\theta_V = {h \over 2 L}\approx 0.014,
\end{align}
where $h=0.64$~cm is the vertical dimension of the nozzle and $L =
23$ cm is the distance between the nozzle and the interaction
region. This means that the vertical position of the atom changes by
only $\approx 3\%$ of the laser beam diameter as the atom crosses
the beam.

A computer code was written in the programming language C to
numerically solve the Louiville density-matrix equations (see for
example Ref. \cite{Cor88}, Ch. 15). The rotating-wave approximation
was applied, and the equations were solved for the upper-state
population using the semi-implicit Euler method \cite{press99}. The
program calculated the spectral line shape by integrating the
density-matrix equations in time for atoms with a given transverse
atomic-beam velocity $v_y$, longitudinal atomic-beam velocity $v_x$,
initial $z_0$ position, and initial $y_0$ position. The program then
integrated over each of these parameters with the proper
distribution functions to arrive at the complete spectral line shape
\footnote{For simplicity, we neglect correlations between
longitudinal and transverse velocity distributions in the atomic
beam. Estimated errors resulting from this approximations are
included in the error budgets presented in Section \ref{sec
acStarkAnalysis}.}.
%\textbf{Dima questions whether this procedure is kosher, and
%wonders what is the size of the correction. Simon is writing a new
%code... Jason: I think we can simply state that we neglect
%correlations between the longitudinal and transverse velocity
%distributions for simplicity.  The analysis with the v=30 cm/ms data
%indicates the type of uncertainties that arise from modifying the
%velocity distribution and any correlation effect is likely to be
%less significant than the uncertainties in the longitudinal velocity
%distribution. }

The upper-state population was integrated from a time $t_i = - 3 \,
\sqrt{2}\, r_0/v_x$ to $t_f = + 3 \, \sqrt{2} \, r_0/v_x$, where
$r_0$ is the $e^{-1}$ electric-field radius.
% and $v_l$ is the longitudinal atomic velocity. \textbf{Dima says: why introduce $v_l$
%when we already have $v_x$?}
The time integral was extended after
$t_f$ to include an exponential decay with a decay time given by the
lifetime of the \TDOne state. The integration over the $z_0$
position, or vertical position of the atom relative to the laser
beam, was performed from $z_i =0$ to $z_f=3 \, \sqrt{2} \, r_0$.
Calculations using these integration ranges were found to give
consistent results with calculations with larger integration ranges.

The integration over the initial $y_0$ position was done by
replacing the $2 \, \pi y_0 / \lambda$ factor in the standing wave
by a phase.  Since the intensity of the standing wave repeats every
$\pi /2$, the integration in this phase was from $0$ to $\pi / 2$.
Atoms traveling at large velocity $v_y$ see many periods of the
standing wave as they traverse the laser beam. For this reason, the
initial phase has little effect on the resulting signal. Thus for
atoms with a magnitude of $|v_y|$ larger than a certain maximum
value, the phase was set to 0 and the signal was not integrated over
the phase. The magnitude of the maximum velocity was empirically
chosen to be $1.5 \times \Gamma_0 \times \lambda/(2\pi)$, where
$\Gamma_0$ is the natural linewidth of the transition.

The integration over the longitudinal velocity assumed a
distribution given for atoms escaping from a hole \cite{ramsey56}
\begin{align}
N(v)={2 \, v^3 \over v_P^4} e^{-v^2 \over v_P^2}, \label{eq
longVDist}
\end{align}
where $v_P = \sqrt{2\, k\, T/m} \approx 2.7 \times 10^4\: {\rm
cm/s}$ is the most probable velocity.

The transverse velocity distribution along $\hat{y}$ used in the
calculation was determined experimentally.  The spectral line shape
of the \SSZeroToTDOne transition was recorded without the
power-build-up cavity. This empirically determined line shape was
then used as the input for the calculation.

A significant amount of effort was put into optimizing the program
so that it was realistic to perform numerous calculations with
different ac-Stark shift parameters and electric-field amplitudes.
The time it took to perform a calculation varied depending on the
values of $\alpha$ and $\varepsilon$.  For most of the calculations
done in this work the time was $20-60$ min. for a single line shape
consisting of 500 frequency points, running on a 3.6 GHz Pentium 4
processor.

\subsection{Calculated Line Shapes}

Experimental data were taken at a variety of different light powers
and dc-electric fields.  The maximum electric field of the standing
wave was varied between 1.5 and 7.5~kV/cm, and the dc electric field
was varied between 5 and 15~kV/cm.  Over this range of field values
the line shape changes quite significantly.

Figure \ref{fig E5CalcLineShape} shows the calculated line shapes
for four different light powers (corresponding to the amplitudes of
the light electric field $\varepsilon_0$ shown in the figure), with
a dc electric field of 5~kV/cm. The ac-Stark parameter assumed for
these calculations is that estimated for the ground-state shift,
$\alpha=0.28 \: {\rm Hz/(V/cm)^2}$. Figure \ref{fig
E15CalcLineShape} shows the line-shape variation with power for the
same ac-Stark-shift parameter, but a dc electric field of 15~kV/cm.
In both figures, the calculated signals have been divided by the
square of the maximum electric field of the light.

\begin{figure}
\centerline{\includegraphics[width=3. in]{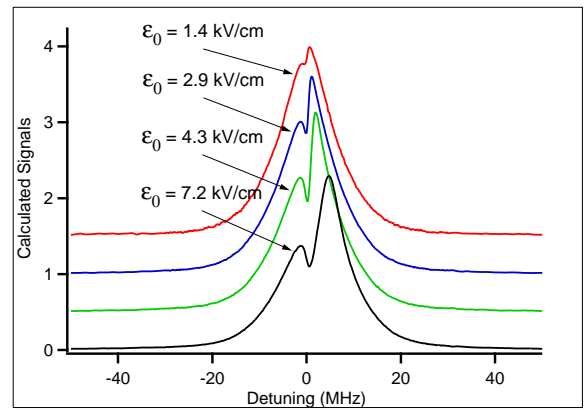}}
\caption[Calculated line shapes for different laser fields for E = 5
kV/cm.]{Calculated line shapes for different laser fields for E = 5
kV/cm.  The ac-Stark shift used in the calculation corresponds to
the estimated ground-state shift. The calculated signals have been
divided by the square of the maximum light electric field.  The
curves are vertically offset for clarity.} \label{fig
E5CalcLineShape}
\end{figure}

While the details of the line shapes are quite complicated, the
basic shape can be understood qualitatively by considering several
different effects. The first effect we discuss is the saturation of
the transition. Despite the smallness of the transition amplitude
($\approx 3 \times 10^{-4}\: {\rm e\, a_0}$ at $15 \: {\rm kV/cm}$),
the intense standing wave present at the highest light powers used
in the experiment is sufficient to saturate the transition.  This is
clearly observed in the decrease of the power-normalized signal size
as the power is increased. The effect is more pronounced for the
calculations at a dc electric field of 15~kV/cm than for those at
5~kV/cm due to the increased Stark-induced amplitude at the higher
electric field.
\begin{figure}
\centerline{\includegraphics[width=3. in]{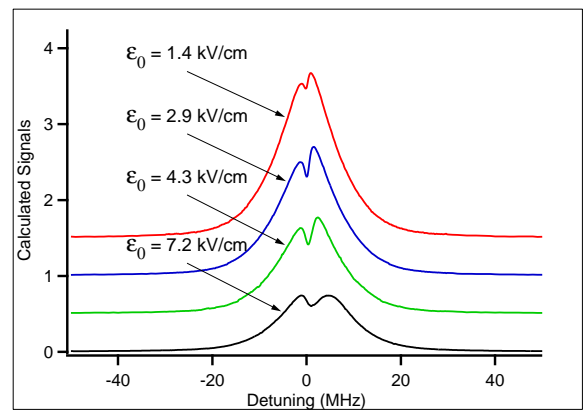}}
\caption[ Calculated line shapes for different laser fields for E =
15 kV/cm.]{Calculated line shapes for different laser fields for E =
15 kV/cm. The ac-Stark shift used in the calculation corresponds to
the estimated ground-state shift. The calculated signals have been
divided by the square of the maximum light electric field. The
curves are vertically offset for clarity.} \label{fig
E15CalcLineShape}
\end{figure}

In addition to the overall decrease in the signal size, saturation
also leads to hole-burning in the velocity distribution. The width
of the dip is comparable to the homogeneous line width, which, in
this case, is the power-broadened width.

%\begin{figure}
%\centerline{\includegraphics[width=5.25 in]{HoleBurning.eps}}
%\caption[Calculated line shapes with no ac-Stark shift for different
%laser fields for E = 15 kV/cm.]{Calculated line shapes with no
%ac-Stark shift for different laser fields for E = 15 kV/cm The
%calculated line shapes have been divided by the square of the
%maximum light electric field.} \label{fig holeBurning}
%\end{figure}

While hole-burning changes the spectral line shape, it does not make
the line shape asymmetric around zero detuning.  To understand the
asymmetry of the line shape we must include the effects of the
ac-Stark shift.  As can be seen from Figs. \ref{fig E5CalcLineShape}
and \ref{fig E15CalcLineShape}, the effect of the ac-Stark shift is
not uniform over the profile of the line, suggesting a dependence on
the atoms' transverse velocity.

The source of the asymmetry can be understood in terms of an
apparent frequency modulation resulting from the ac-Stark shifts. As
the atoms travel through the laser beam the resonance frequency of
the transition shifts depending on the intensity of the local light
field. From an atom's perspective this is equivalent to a frequency
modulation of the laser light.

The depth of the frequency modulation (i.e., the maximum frequency
shift) is determined by the magnitude of the ac-Stark shift and is
given by
\begin{align}
\xi = {1 \over 2} \, \alpha \, \varepsilon_0^2.
\end{align}
With the estimated ground-state shift and light powers used in the
experiment, the depth of the modulation ranges between 0.28 and 7.4
MHz. The frequency of the modulation (as the atom moves between
peaks and valleys of the standing light wave) is determined by the
transverse velocity
\begin{align}
\frac{\Omega}{2\pi} = {2 \, v_y \over \lambda} = 2 \Delta \nu_D,
\label{eq modFreq}
\end{align}
where $\Delta \nu_D$ is the Doppler shift of an atom moving with
transverse velocity $v_y$. The factor of two in the above equation
arises from the fact that the ac-Stark shift is quadratic so that
the period of the shift is half that of the wave length of the
light.

Since the apparent frequency of the modulation is determined by the
transverse velocity [Eq. \eqref{eq modFreq}], the effect is quite
different for atoms with different $v_y$. The transverse velocities
where the character of the frequency modulation changes, depend on
the size of the ac-Stark shift, and, as discussed below, on the
homogeneous line width of the transition.

\begin{figure}
\centerline{\includegraphics[width=3. in]{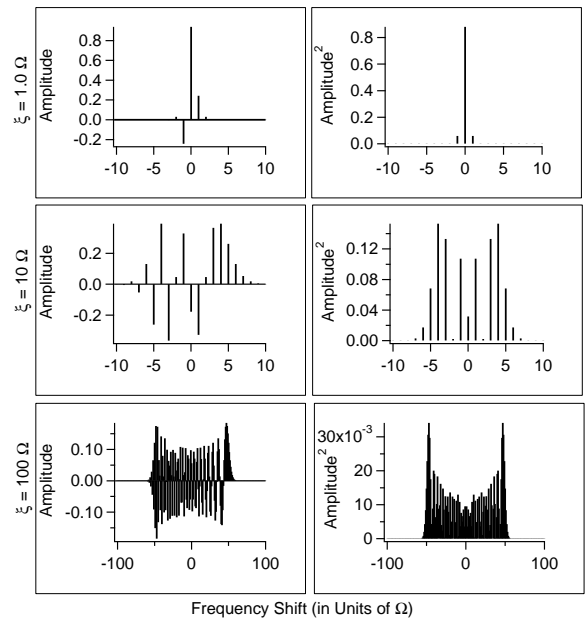}}
\caption[Amplitude and intensity spectra for frequency modulated
light with different modulation indices.]{Amplitude and intensity
spectra for frequency modulated light with different depths of
modulation.  The amplitudes are given by the values of the
appropriate Bessel functions (see, for example, Ref.
\cite{Arf2005}).} \label{fig FreqMod}
\end{figure}

The character of frequency modulation is determined by the relative
size of the depth of the modulation and the frequency of the
modulation.  In the case where $\xi/2 << \Omega$, frequency
modulation is characterized by the appearance in the spectrum of the
light-frequency detuning as perceived by the moving atom of
principal sidebands located at $\pm \Omega$ of the fundamental
frequency.  The sign of the amplitude of the sidebands is opposite
for the high- and low-frequency sidebands. As the depth of the
modulation is increased, the sidebands increase and higher-order
sidebands develop. Eventually, the bulk of the amplitude shifts away
from the fundamental frequency to the extreme frequencies.  This
evolution is shown in Fig. \ref{fig FreqMod} for the electric-field
amplitude and intensity of the light (proportional to the square of
the amplitude).

An atom with a transverse velocity $v_y = \Delta \nu_D \, \lambda$
sees two ``carrier" frequencies,
\begin{align}
\nu_1&=\nu_0 - \Delta \nu_D \\
\nu_2&=\nu_0 + \Delta \nu_D,
\end{align}
corresponding to the Doppler shifts of the two counter-propagating
waves which make up the standing wave.  Since the resonant frequency
of the transition is modulated, the detuning of the atomic resonance
relative to the light frequency acquires sidebands.  From the atom's
perspective, this appears as sidebands spaced a distance $2 \Delta
\nu_D$ around $\nu_1$ and $\nu_2$. Thus, the high-frequency side
band of $\nu_1$ overlaps with $\nu_2$ and the low-frequency side
band of $\nu_2$ overlaps with $\nu_1$. However, since the
high-frequency sideband for $\nu_1$ is positive, while the
low-frequency sideband of $\nu_2$ is negative, there is a reduction
of the amplitude at $\nu_1$ and an enhancement at $\nu_2$. This
leads to asymmetries in the resonant peaks at $\pm \Delta \nu_D$. As
the transverse velocity is increased, the amplitude of the sidebands
decreases and the height of the two Doppler-shifted resonances evens
out.

Atoms with small transverse velocity pass through the laser beam
without traversing multiple nodes and anti-nodes of the standing
wave. The nature of the modulation is therefore not sinusoidal for
these atoms, and cannot be parameterized in terms of $\Omega$.  The
atoms experiencing this regime have transverse velocities
corresponding to Doppler shifts less than the transit line width
which is $\approx 200$ kHz. Furthermore, if $\Omega$ is less than
the homogeneous line width (determined in this case by the
power-broadened width), then the effects of the frequency modulation
are smeared out.  For these reasons, the regime of harmonic
modulation where $\Omega$ is much smaller than $\xi/2$ is never
realized in the system under consideration.

Even if the atoms move exactly perpendicular to the light, there is
spectral asymmetry due to the spatial dependence of the ac-Stark
shift -- the shift is different depending on the phase of the light
wave the atoms see (determined by the y coordinate), and on the
position of the atoms in the x-z plane.

The resulting spectrum is the sum of all velocity classes, this sum
leads to asymmetries near the center of the spectra shown in Figs.
\ref{fig E5CalcLineShape} and \ref{fig E15CalcLineShape}.

%For atoms that have large transverse velocities, so that $\Omega>
%\xi/2$, the atom sees frequency modulation in the regime where the
%modulation frequency is larger than the depth of the modulation and
%the modulation is characterized by sidebands spaced a frequency $\pm
%\Omega$ around the main frequency.

\section{Analysis and Results}
\label{sec acStarkAnalysis}

Line shapes of the type shown in Figs. \ref{fig E5CalcLineShape} and
\ref{fig E15CalcLineShape} were generated over the appropriate range
of transition amplitudes and a range of ac-Stark-shift parameters.
The calculated line shapes were used to create an interpolating
function in frequency, the ac-Stark-shift-parameter, and transition
amplitude.

The calculations were related to the experimental spectral line
shapes using the following parameterization of the calculations
\begin{align}
\mathcal{L}(\nu) = a \, f\prn{\varepsilon_0, \, \alpha, \, \beta, \,
E, \, \theta, \, s \prn{\nu - \nu_c}}, \label{eq fitEq}
\end{align}
where $a$ is the amplitude of the signal, and the function
$f\prn{\varepsilon_0, \, \alpha, \, \beta, \, E, \, \theta, \, s
\prn{\nu - \nu_c}}$ is the calculated line shape for a maximum
electric field of the optical standing wave of $\varepsilon_0$,
ac-Stark shift polarizability $\alpha$, Stark-transition
polarizability $\beta$, dc electric field $E$, polarization angle
$\theta$, and frequency $s (\nu - \nu_c)$, centered around $\nu_c$.
The frequency term includes a multiplicative scaling factor $s$.
This $s$ parameter was introduced in order to account for any
variation in the overall width of the data relative to the
calculation.  Such a variation might arise due to a misalignment of
the atomic beam relative to the axis of the power-build-up cavity or
a deviation of the velocity distribution from that measured without
the cavity. We note that the parameters in Eq.\ \eqref{eq fitEq} are
not independent. Aside from scale and offset parameters, the
calculated line shape depends only on the Rabi frequency of the
transition ($\Omega_R = \beta \, E \, \varepsilon_0 \, {\rm
sin}\theta$) and the ac-Stark shift ($\Delta \mathcal{E}  = - 1/2 \:
\alpha \, \varepsilon_0^2 $).

Approximately twenty experimental scans were performed with a given
dc-electric field, polarization angle, and light power.  These
$\approx 20$ scans were grouped together to form a set of data which
was analyzed as described below. The dc-electric field was varied
from 5 to 15 kV/cm between sets. Polarization angles of $90^\circ$
and $\pm 45^\circ$ relative to the dc-electric field were used.  The
light power was varied from $180~\mu{\rm W}$ to $2.4~{\rm mW}$
transmitted (corresponding to a maximum light electric field inside
the cavity of 2 to 7.4 kV/cm). Data were taken at a total of 45
different field parameters.

The maximum light electric field inside the cavity, $\varepsilon_0$,
was determined using the measured power transmitted through the
cavity and the experimentally determined transmission coefficient of
the cavity mirrors [Eq.\ \eqref{eq mirrorTrans}]. The dc-electric
field was measured using the voltage divider mentioned in Section
\ref{ssec efield}.  The value of the Stark transition polarizability
was fixed in the analysis at $\beta = 2.18 \times 10^{-8} \: e\, a_0
/({\rm V/cm})$ [see Eq.\ \eqref{eq beta}]. Effects arising from the
uncertainty in $\beta$, as well as an independent evaluation of
$\beta$ from the present data are discussed below.

\begin{figure}
\centerline{\includegraphics[width=3.
in]{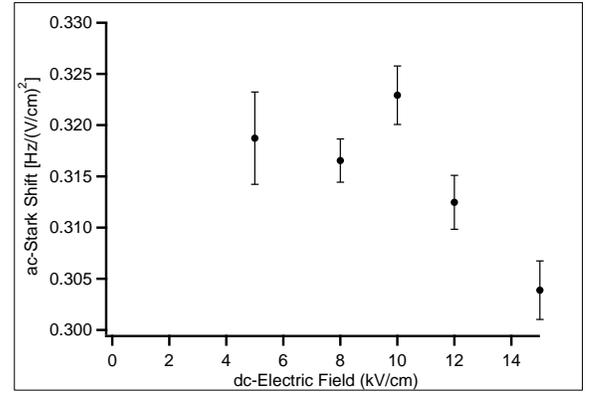}}
%\centerline{\includegraphics[width=3. in]{dcFieldDepB218.eps}}
\caption{The magnitude of the ac-Stark shift parameter as a function
of dc electric field. The value of the Stark transition
polarizability assumed in this analysis is $\beta = 2.18 \:{\rm e\,
a_0/(V/cm)}$. The error bars in this plot are based on the spread of
the extracted ac-Stark parameters for a given value of the dc
electric field.} \label{fig dcFieldDep}
\end{figure}

The power-normalized scans were individually fit to Eq.\ \eqref{eq
fitEq}. Typical scans and fits for three different combinations of
dc electric field and laser light power are shown in Fig.\ \ref{fig
scanFits}. For a given power and dc electric field the mean fit
values of a set of scans were determined, and the distribution of
the values was used to estimate the statistical error. The overall
result, including sets at all of the dc-electric fields,
polarization angles, and light powers used in the experiment is
\begin{align}
\alpha_0^{ac}(^3{\rm D}_1) + \alpha_2^{ac}(^3{\rm D}_1)& -
\alpha_0^{ac}(^1{\rm S}_0) \nonumber \\  & = -0.3123(11) \:{\rm
Hz/(V/cm)^2},
\end{align}
where the error is determined from the statistical errors in each
set of scans.  The reduced chi squared for the data is 3.18 for 45
points. The large reduced chi squared is primarily due to
inconsistency in the data taken at a dc electric field of 15~kV/cm
with the data taken at lower fields. Table \ref{tab dcFieldValues}
shows the results for the different dc electric fields and Fig.\
\ref{fig dcFieldDep} shows the dc-field dependence of the ac-Stark
shift parameter.

\begin{figure}
\centerline{\includegraphics[width=3. in]{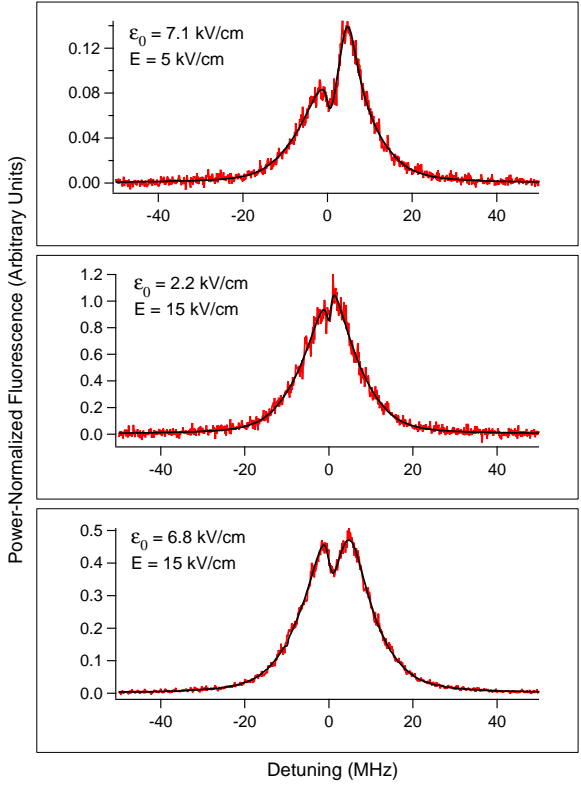}}
\caption{Experimental power-normalized line shapes and fits to the
model described in the text for three different combinations of dc
and light electric field values. Typical time for each scan was 200
ms. For the $\varepsilon_0=6.8\ $kV/cm (2 mW of transmitted light
power), E=15 kV/cm data, the peak detected PMT-anode current was
$\approx1.2\ \mu$A.} \label{fig scanFits}
\end{figure}

\begin{table}
\begin{center}
\begin{tabular}{lcccc} \hline \hline
dc Electric Field & ac-Stark Shift & $\chi^2$ & N \\
\hline
5 kV/cm & -0.3187(51) & 0.77 & 4 \\
8 kV/cm & -0.3165(31) & 0.46 & 8 \\
10 kV/cm & -0.3229(20) & 1.99 & 16 \\
$\; \; \;$ Pol. $= 0^\circ$ & -0.3170(46) & - & 1 \\
$\; \; \;$ Pol. $= 45^\circ$ & -0.3211(31) & 0.973 & 10 \\
$\; \; \;$ Pol. $= -45^\circ$ & -0.3282(33) & 4.12 & 5 \\
12 kV/cm & -0.3125(29) & 0.82 & 2 \\
15 kV/cm & -0.3039(16) & 3.21 & 15 \\ \hline \hline
\end{tabular}
\end{center}
\caption{Extracted values of ac-Stark shift for different dc
electric fields and angles of light polarization with respect to the
dc electric field. The value of the Stark transition polarizability
assumed in this analysis is $\beta = 2.18 \; {\rm e\, a_0/(V/cm)}$.
The units for the ac-Stark shift are ${\rm Hz/(V/cm)^2}$. \label{tab
dcFieldValues}}
\end{table}

Excluding the data taken at 15 kV/cm gives a value of the ac-Stark
shift of
\begin{align}
\alpha_0^{ac}(^3{\rm D}_1) + \alpha_2^{ac}(^3{\rm D}_1) & -
\alpha_0^{ac}(^1{\rm S}_0) \nonumber
\\ & = -0.3188(14) \:{\rm Hz/(V/cm)^2},
\end{align}
with the reduced chi squared of 1.57 for 30 sets.  Excluding
everything but the 15 kV/cm data gives
\begin{align}
\alpha_0^{ac}(^3{\rm D}_1) + \alpha_2^{ac}(^3{\rm D}_1) & -
\alpha_0^{ac}(^1{\rm S}_0) \nonumber
\\ & = -0.3039(16) \:{\rm Hz/(V/cm)^2},
\end{align}
with a reduced chi squared of 3.21 for 15 sets.

\begin{table}
\begin{center}
\begin{tabular}{lcccc} \hline \hline
Analysis Conditions & ac-Stark Shift &  $\chi^2$ & N  \\ \hline
$\beta=2.18\times 10^{-8}\: {\rm e\,
a_0/(V/cm)}$ & -0.3123(11) & 3.18 & 45 \\
$ \; \; \; $ Excluding 15 kV/cm data & -0.3188(14) &  1.57 & 30 \\
$ \; \; \; $ Only 15 kV/cm data & -0.3039(16) &  3.21 & 15 \\
$\beta=2.08\times 10^{-8}\: {\rm e\,
a_0/(V/cm)}$ & -0.2953(11) & 7.53 & 45 \\
$ \; \; \; $ Excluding 15 kV/cm data & -0.3112(14) &  1.88 & 30 \\
$ \; \; \; $ Only 15 kV/cm data & -0.2791(14) &  1.93 & 15 \\
$\beta=2.28\times 10^{-8}\: {\rm e\,
a_0/(V/cm)}$ & -0.3290(11) & 2.46 & 45 \\
$ \; \; \; $ Excluding 15 kV/cm data & -0.3292(14) &  1.92 & 30 \\
$ \; \; \; $ Only 15 kV/cm data & -0.3286(19)& 3.76 & 15 \\
Misaligned Velocity Distribution; & & &\\
$\; \; \; \beta = 2.18\times
10^{-8}\: {\rm e\, a_0/(V/cm)}$ & -0.3076(11) & 10.23 & 45\\
$\; \; \; \beta = 2.08\times
10^{-8}\: {\rm e\, a_0/(V/cm)}$ & -0.2930(10) & 18.46 & 45\\
$\; \; \; \beta = 2.28\times
10^{-8}\: {\rm e\, a_0/(V/cm)}$ & -0.3259(12) & 5.25 & 45\\
`s' Parameter fixed at unity & -0.3384(10) & 9.16 & 27 \\
Atomic velocity $v_P= 30\, {\rm cm/ms}$ & -0.2963(11) & 7.77 & 45 \\
 \hline \hline
\end{tabular}
\end{center}
\caption{Extracted values of ac-Stark shift for different analysis
procedures and assumptions.  The units of the ac-Stark shift are
${\rm Hz/(V/cm)^2}$. For each case, the value of the reduced
$\chi^2$, and the number of sets N used in the analysis are
presented. For all cases except the one presented in the last line,
the most probable velocity in the beam is assumed to be $v_P = 27 \:
{\rm cm/ms}$.} \label{tab alphaResults}
\end{table}

The discrepancy is also apparent in the $s$ parameter described
above.  The $s$ parameter has a clear dependence on the dc-electric
field, going from $s=0.984(3)$ at 5 kV/cm to $s=0.9535(6)$ at 15
kV/cm. A value of the $s$ parameter less than unity corresponds to
an experimental line shape which is broader than the line shape
calculated assuming $s=1$.

This broadening of the line shape might suggest that the
inconsistency in the data is a result of not properly accounting for
saturation effects. However, there is no clear dependence of the
extracted ac-Stark shift value on the light power, which one would
expect if the saturation effects were the source of the discrepancy.

We analyzed the data in a number of different ways in order to
understand this discrepancy as well as estimate systematic errors
from calculational assumptions.  The results of these investigations
are tabulated in Tab.\ \ref{tab alphaResults} and are discussed
below.

\begin{figure}
\centerline{\includegraphics[width=3. in]{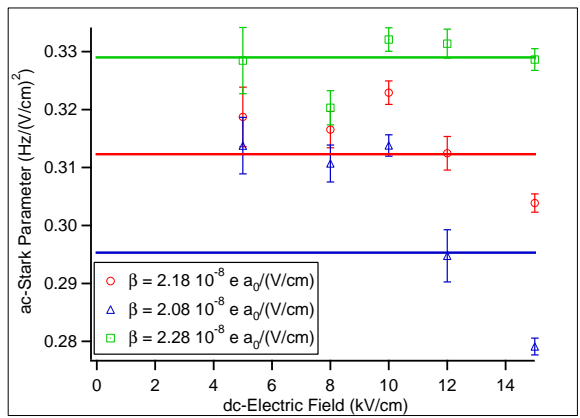}}
\caption{The magnitude of the ac-Stark-shift parameter as a function
of dc electric field for different values of the transition
polarizability $\beta$.} \label{fig dcFieldDepAllBetas}
\end{figure}

The ac-Stark shift and the value of the Stark transition
polarizability, $\beta$, are highly correlated in the
parametrization of the spectral line shape.  The effects of the
uncertainty in the value of $\beta$ were investigated by reanalyzing
the data with the procedure described above assuming different
values of $\beta$.  The results of this analysis are shown in Tab.\
\ref{tab alphaResults}. Figure \ref{fig dcFieldDepAllBetas} shows
the dependence of the ac-Stark shift parameter on the dc electric
field for the different values of the Stark transition
polarizability. This analysis reveals a definite dependence of the
ac-Stark shift parameter determined from the analysis on the value
of $\beta$. Furthermore, the discrepancy between the 15~kV/cm data
and the lower field data was reduced when the data were analyzed
with a value of the Stark transition polarizability of $\beta = 2.28
\times 10^{-8}\: {\rm e\, a_0/(V/cm)}$. However, the data are still
inconsistent with reduced chi-squared of 2.46 for 45 points.
%\textbf{D: I wonder what is the internal consistency of the 8, 10,
%12, and 15 kV/cm data. It might be if the errors are properly
%assigned, the inconsistency would go away. J: see TAble I}
This analysis suggests that the value of the Stark transition
polarizability may be higher than the central value in Eq.\
\eqref{eq beta}, although consistent within the quoted uncertainty.
%Given the remaining lack of consistency in
%the data, we do not suggest that this analysis provides a more
%accurate determination of the value of the Stark transition
%polarizability. \textbf{D: I would not make the preceding statement
%just yet, we can do it later after we attempt the extraction. J: We
%just finish showing that the reduced chi squared decreases and the
%data is more consistent with the larger beta.  I think we need to
%make some statement as to what we infer from this concerning the
%value of beta}
Below, we describe a procedure for determining the value of the
Stark transition polarizability from these data. This analysis
provides a determination of the uncertainty in the ac-Stark shift
parameter resulting from the uncertainty in $\beta$. Based on the
results presented in Tab.\ \ref{tab alphaResults}, we arrive at an
uncertainty in the ac-Stark shift due to uncertainty in $\beta$ of
$5 \%$.
%\textbf{D: From Fig. 9, it looks like it should be more like
%10\% to me. Also, analysis at $\beta=2.38$ woyld be very interesting
%to see, but I am not suggesting it at this point because this is so
%labor intensive.}

Another possible source of experimental error arises from
uncertainties in the atomic velocity distribution along the axis of
the power build-up cavity.  This uncertainty is the motivation for
introducing the $s$ parameter described above and is highly
correlated with it.  The fact that the $s$ parameter is less than
unity suggests that the center velocity of the atomic beam is not
exactly perpendicular to the cavity axis.  To analyze the effects of
such a misalignment, the calculation described above was modified to
include a Doppler distribution which was misaligned from the axis of
the power-build-up cavity by $0.14^\circ$. This misalignment
corresponds to a Doppler shift of the two counter-propagating light
waves constituting the standing wave by $\pm$2~MHz, respectively, as
seen from the point of view of the atoms in the atomic beam moving
with the most probable velocity. Misalignments larger than this lead
to noticeable effects in the velocity distribution which were not
present in the experimental data. We believe this to be an upper
limit on the possible misalignment. The data were analyzed as
before; the results are presented in Tab.\ \ref{tab alphaResults}.
From these we determine that the uncertainty in the ac-Stark shift
parameter due to the uncertainties in the alignment of the atomic
beam with the cavity axis are $2 \%$. The values of the $s$
parameter resulting from this analysis ranged from 1.05 for E=5
kV/cm to 1.01 for E=15 kV/cm.
%The fact that the $s$ parameter lies
%above unity (meaning the experimental line shape is narrower than
%the the \textbf{line shape calculated assuming $s=1$}) further
%support our belief that the misalignment used in the calculation
%represents an upper limit for possible misalignment present in the
%experiment. \textbf{DB: This place keeps giving me doubts. What if
%there was a misalignment when the measurement without the cavity was
%done ? }
% \textbf{because the
%$s$ parameter was always less than unity in the analysis that did
%not assume the beam misalignment -- this is what is apparently
%assumed, but this does not sound very convincing. J: If we believe
%that we know the velocity distribution (from the measurement without
%the cavity) then any misalignment of the beam relative to the cavity
%broadens the experimental lineshape relative to the calculation and
%shows up in the 's' parameter.  The calculations with the 2 MHz
%misalignment are broader than the data for all dc-field values.  If
%the calculation is getting the wings of the distribution correct
%(for ANY of the dc-electric fields), then the velocity distribution
%cannot be misaligned relative to the cavity axis by 2 MHz.}

We also investigated the effect of including the $s$ parameter in
the fit.  We analyzed a fraction of the data (27 sets) fixing the
$s$ parameter to unity. The fits using this procedure were not as
good as the fits where $s$ was allowed to vary. In this case, the
values of the ac-Stark shift parameter are largely determined by the
wings of the Doppler distribution, since that is the only parameter
available to broaden the line shape.  The result of this analysis is
shown in Tab. \ref{tab alphaResults}.

The longitudinal velocity distribution used in the calculation is
also a possible source of uncertainty.  We assume a standard
longitudinal velocity distribution of effusive atoms escaping from a
hole of [Eq.\ \eqref{eq longVDist}], with the temperature fixed by
the measured temperature at the rear (i.e., the coldest part) of the
oven. However, atomic beams frequently deviate from this textbook
distribution due to preferential scattering of the slow atoms (the
so-called Esterman effect, see, for example, a discussion in Ref.
\cite{Reg2001}). In addition, the oven is not at a uniform
temperature; the temperature varies by over $100 \: ^\circ {\rm C}$
from front to back of the oven. This variation in temperature can
modify the observed velocity distribution of the atomic beam. In
order to investigate the effects of uncertainty in the longitudinal
velocity distribution, data were analyzed using calculations
assuming velocity distributions corresponding to different
temperatures.  The extracted values of the ac-Stark shift parameter
are shown in Tab.\ \ref{tab alphaResults}. Assuming a maximal
deviation in the central temperature of $100 ^\circ$, we arrive at
an uncertainty in the ac-Stark shift parameter of about $5 \%$.

\begin{figure}
\centerline{\includegraphics[width=3. in]{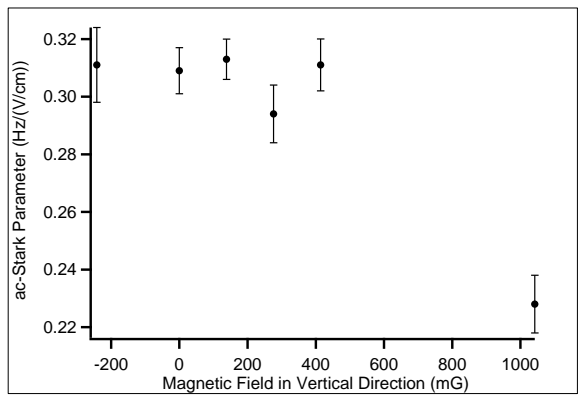}}
\caption{Value of the extracted ac-Stark shift parameter as a
function of magnetic field in the vertical direction.  The dc
electric field for these data ranged between 5 kV/cm and 15 kV/cm.}
\label{fig bFieldDep}
\end{figure}

Another key assumption in the calculation is that the problem can be
reduced to that involving just two atomic levels.  If the dc
electric field and light field are the only fields present and are
aligned appropriately, this is a correct assumption (see Sect.\
\ref{sect calc}). However, if there are additional fields present,
such as a magnetic field, this assumption may break down.  With the
experimental geometry used here (Fig.\ \ref{fig_Exp_Arr}) the strong
dc electric applied along $\hat{z}$ splits the $M_z=\pm 1$
$^3\rm{D}_1$ states from the $M_z=0$ state. A residual field along
the $x$- or $y$-axes, would lead to mixing between the $M_z=\pm 1$
and $M_z=0$ states. However, since the Zeeman shifts ($\Delta \nu =
g \, \mu \, B \approx 35 \; {\rm kHz}$) corresponding to the
residual magnetic fields are small relative to the dc-Stark shifts,
($\Delta \nu = 1/2 \: \alpha_2 \, E^2 \approx 100 - 800 \: {\rm
kHz}$ \cite{bowers99}),
this mixing is suppressed due to the strong dc electric field along
the $z$-axis and the two-level approximation still holds
\footnote{Additionally, if this effect were to be the problem, one
would expect larger internal inconsistencies in the data taken at
low rather than at high dc electric field where the dc-Stark shifts
are small. This is opposite to what we actually observed.}. However,
a residual magnetic field along the $z$-axis would serve to split
the degenerate $M=\pm1$ states and the two-level treatment would be
incorrect. To investigate this possibility, we utilized the
magnetic-field coil normally used to compensate the ambient field in
the $z$-direction to apply a magnetic field parallel to the dc
electric field.  Data were then taken at different values of
magnetic fields, which were analyzed as described above.  The
results are shown in Fig.\ \ref{fig bFieldDep}.  At large values of
the magnetic field ($\approx 1$ G), there is a dependence in the
extracted value of the ac-Stark shift parameter as expected.
%\textbf{D: This needs explanation. Specifically, at what magnetic
%field do we expect to see a deviation? I think the parameter may be
%something like Zeeman shift over natural or transit width (in other
%words, the width of the feature). Above we were comparing Zeeman
%shift and dc-Stark shift, so this is a bit confusing.}.
However, no
dependence was observed for data taken at magnetic field values that
might realistically be present in the experiment. Thus, using the
errors for the data taken with varying magnetic field we estimate
the uncertainty in the ac-Stark shift parameter due to residual
magnetic-field effects to be $<3 \%$.
%\textbf{D: Need to explain
%explicitly what this number is based on. If this from the low-field
%part of Fig. 10, wouldn't one need to know where the curve actually
%turns over? Why does the negative-B point have large error bar ? J:
%Comment of B-field dependence on pg 13:  Error was very
%conservatively estimated based on the error bars on the points,
%which are around 3\%. This was less than other errors so I didn't
%bother pushing it further.  A better way is to fit the points to a
%line. You get a slope which is consistent with zero. You can use
%that error on that slope and the fact that we think the error in
%$B_z$ is 50 mG to estimate the deviation in alpha.  This leads to a
%0.3\% error.  This assumes a linear dependence of alpha on B and it
%simply isn't important to quantify it at that level. The points on
%the plot are combinations of different electric fields and different
%numbers of data sets. The point at negative magnetic field has
%larger error bars because it has different combinations of fields
%and data sets. The point of running at different fields was to
%experimentally look for an effect since we didn't have a model. We
%knew that the at some level the magnetic field would affect things.
%We successfully determined that level was above the 50 mG we think
%we had.  I don't know where the effect starts to be important; we
%didn't investigate it that carefully and we don't need to in order
%to determine the effect isn't important for this measurement.}

\begin{figure}
\centerline{\includegraphics[width=3. in]{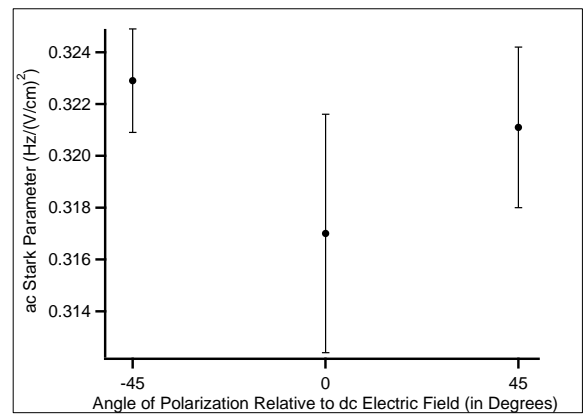}}
\caption{Values of the extracted ac-Stark-shift parameter as a
function of polarization angle.  The dc electric field was 10 kV/cm
for these data.} \label{fig polDep}
\end{figure}

Misalignment of the optical field relative to the dc-electric field
can also lead to a splitting of the energy eigenstates that are
excited. With the geometry used in this experiment (Fig.\
\ref{fig_Exp_Arr}), there are two possible kinds of such
misalignment. One is an error in the angle between light field and
dc electric field. Since the ac-Stark shift does not depend on the
angle of the light field relative to the dc field in this geometry
(see Sect.\ \ref{ssec acStarkOnStark}), a misalignment of this kind
does not lead to a breakdown of the two-level approximation, but
changes the transition matrix element [Eq.\ \eqref{eq starkAmp}]. In
the other kind of misalignment, the cavity axis has a component
along the dc electric field. In this case, the assumption of a
two-level system generally breaks down. However, for the case where
the light field is polarized along the $x$-axis, the dc electric
field and the light field remain perpendicular. Thus, the above
argument that was used to reduce the problem to that involving just
two atomic levels is valid in this case, and the misalignment does
not affect the extracted value of the ac-Stark-shift parameter. For
a general polarization, a misalignment could influence the ac-Stark
shift measurement. We can place a limit on the uncertainties arising
from such possible misalignment by comparing the extracted
ac-Stark-shift parameters for data taken with the angle of the light
polarization relative to the dc electric field of $\pm45^\circ$ with
data taken at an angle of $0^\circ$. Figure \ref{fig polDep} shows
the value of the ac-Stark shift parameter as a function of the
polarization angle for data taken at a dc-electric field of 10
kV/cm.
%\textbf{D: Is this
%all the data we have for different angles? J: It's all we've ever
%had.}
Using these results we are able to place a limit on the uncertainty
in the ac-Stark-shift parameter due to misalignment of the light-
polarization of $2 \%$.
%\textbf{D: How is this done? I suppose you
%assume a difference in $\alpha$ of something like 0.01 between zero
%and +/-45 degrees. What then do you assume for the dependence of
%systematics on the angle? Here is a Kosher way of doing it (too late
%now): introduce a deliberate known misalignment, and then measure
%the dependence a la Fig. 11 accurately. Then, comparing to a similar
%dependence without deliberate misalignment, you can actually put a
%quantitative limit on the misalignment, and the systematics. This is
%the Commins method of dealing with systematics. We'll need to employ
%this sort of a technique for PNC. J: Your assumption is correct. The
%points agree within errors and the uncertainty in the agreement is
%2\%.  I make no assumption about the form of a possible systematic.}

Additional errors in the data arise from uncertainties in the
dc-electric-field and light-field values.  These errors, along with
those discussed above are listed in Tab.\ \ref{tab errorBudget}. The
uncertainty in polarization alignment was estimated based on the
relative signal sizes at the different polarization angles. The
uncertainty of $\approx 2^\circ$ is comparable to the measured
degree of ellipticity in the transmitted light due to a
birefringence of the cavity mirrors.

\begin{table}
\begin{center}
\begin{tabular}{ccc} \hline
\hline Parameter & Uncertainty & Effect on $\alpha$
\\ \hline
$\beta$ & $5\%$ & $5\%$ \\
Transit Velocity & $5\%$ & $5\%$  \\
Calculational Approximations & n/a & $5\%$ \\
dc Electric Field & $1 \% $ & $1\%$ \\
Polarization alignment & $\approx 2^\circ$ & $2\%$\\
Residual Magnetic Field & $<50 \, {\rm mG}$& $3\%$ \\
Mirror Transmission & $3\%$ & $3\%$ \\
Power & $4\%$ & $ 4\%$\\ \hline  Total Uncertainty & - & $11 \%$
\\\hline \hline
\end{tabular}
\end{center}
\caption{Errors and the resulting uncertainty in the value of the
ac-Stark shift parameter.} \label{tab errorBudget}
\end{table}

Including all of the systematic errors listed in Tab. \ref{tab
errorBudget} we arrive at a systematic error of $11 \%$ in the
ac-Stark shift parameter, while the spread of the data is $\approx
6\%$.
%which is within the error. \textbf{J: I'm not too sure I want to say
%this since many of the errors tabulated in Tab. \ref{tab
%errorBudget} affect all of the data sets in the same way. Do we
%separate out errors like the mirror transmission and power
%calibration which lead to an overall error from transit
%velocity/beta which give different errors?.}  \textbf{D: Before
%deciding on this, we need to discuss what is the source of the
%spread of the data. Somehow, we need to separate our uncertainties
%into systematic and statistical, even if the spread of the data is
%not shot noise. We have to do it as this is SOP. J:
% Every error we have beyond what is quoted in Tab. II and IV is systematic.
%The question is what kind of systematic.  The bottom line is we
%don't understand the spread, but it's clearly not statistical.  The
%systematic errors we do have affect the data in different ways, some
%are calibration and some apparently lead to spread in the values.}
We arrive at a final value of the ac-Stark transition polarizability
of
\begin{align}
\alpha_0^{ac}(^3{\rm D}_1) + \alpha_2^{ac}(^3{\rm D}_1)& -
\alpha_0^{ac}(^1{\rm S}_0) \nonumber \\  & = -0.312(34) \:{\rm
Hz/(V/cm)^2}.\label{Eq_alpha_final}
\end{align}

In an attempt to determine the value of $\abs{\beta}$ from the data,
we also analyzed the results using a global fit of all data. In
addition, the values of the Stark transition polarizability [Eq.
\eqref{eq beta}] and the frequency scaling parameter $s$ directly
affect the extracted value of the ac-Stark polarizabilities.  In
order to properly account for this correlation, the variance of the
data with the calculation was simultaneously minimized with respect
to all three of these parameters.  The data were analyzed by varying
$\alpha$, $\beta$, and $s$ and fitting the individual data scans to
the amplitude, $a$, and the center position $\nu_c$. For a given set
of $\alpha$, $\beta$, and $s$ the $\chi^2$ of the individual fits
were summed. This gave a global ``$\chi^2$ map," using which a
minimum was found with respect to $\alpha$, $\beta$, and $s$.

The value of the minimum of the global $\chi^2$ was normalized to
unity. The errors were assigned by determining the deviations in
$\alpha$, $\beta$, and $s$ required to give a $\chi^2 = 1 + 1 / N$,
where $N$ is the total number of data points used in all of the
fits, and taking the appropriate projections along $\alpha$,
$\beta$, and $s$.  The results for the data grouped by field
configurations as well as the result of a global fit to all of the
data are shown in Tab.\ \ref{tab chiSqdValues}.
%\textbf{D: I propose
%to make plots of these one under the other as a function of the
%electric field to look for trends, correlations, and to compare with
%Fig. 9 J: I think we have entirely too much discussion on the
%different analysis techniques and too many figures as it is.  I
%still think we need to cut this discussion significantly and adding
%a figure isn't going to help. D: What is the motivation for cutting
%the paper? Also, at this point, I feel we are still working on the
%subject matter, at least to a degree. I'll plot these things up to
%see if this jolts some thoughts.}

\begin{table}
\begin{center}
\begin{tabular}{lcccc} \hline \hline
dc Electric Field & ac-Stark Shift  & $\beta$ & s & N \\
\hline
5 kV/cm & -0.3261(38) & 2.301(18) & 0.9931(26) & 4 \\
8 kV/cm & -0.3196(11) & 2.166(5) & 0.9724(8) &8 \\
10 kV/cm &  &  & &  \\
$\;\;\; {\rm Pol.} =0^\circ$ & -0.3289(11) & 2.235(3) &0.9788(21) & 1 \\
$\;\;\; {\rm Pol.} =\pm 45^\circ$ & -0.3327(9) & 2.283(8) & 0.9790(6) & 15 \\
12 kV/cm & -0.3248(13) & 2.233(4) & 0.9654(9) & 2 \\
15 kV/cm & -0.3094(13) & 2.203(5) & 0.9622(5) & 15 \\
All Data & -0.3284(5) & 2.237(2) & 0.9730(3) & 45 \\
\hline \hline
\end{tabular}
\end{center}
\caption{Extracted values of ac-Stark shift, Stark transition
polarizability, and `s' parameter as determined by chi-squared fits
to the indicated data sets. The units of the ac-Stark shift are
${\rm Hz/(V/cm)^2}$ and the units of $\beta$ are $10^{-8} \, {\rm
e\, a_0/(V/cm)}$.\label{tab chiSqdValues}}
\end{table}
%
%\begin{figure}
%\centerline{\includegraphics[width=3.
%in]{GlobalFitResults.eps}}\caption{\textbf{Global fit. Need to refer
%to this in text}}\label{fig Global_Fit}
%\end{figure}

There is good agreement between the results determined from the
chi-squared map and the result determined from the fits to the
individual data sets.  Again, there is a discrepancy between the
different electric-field sets. In addition, the extracted value of
the Stark transition polarizability $\beta$ is inconsistent among
the different electric field sets. Using the spread as an estimate
of the uncertainty, we arrive at a measurement of the Stark
transition polarizability of
\begin{align}
\abs{\beta} = 2.24^{+0.07}_{-0.12} \times 10^{-8}\, {\rm e \,
a_0}/{\rm V/cm},\label{Eq_beta_final}
\end{align}
which is in good agreement with and of comparable uncertainty as the
previous measurement [Eq.\ \eqref{eq beta}]. The error budget for
the determination of $\beta$ is presented in Table \ref{tab
errorBudget_beta}. Considering the two measurements independent, we
get the final value
\begin{align}
\abs{\beta} = 2.19(8) \times 10^{-8}\, {\rm e \, a_0}/{\rm
V/cm}.\label{Eq_beta_final_final}
\end{align}

\begin{table}
\begin{center}
\begin{tabular}{ccc} \hline
\hline Parameter & Uncertainty & Effect on $\beta$
\\ \hline
$E$ & $1\%$ & $1\%$ \\
Polarization alignment & $\approx 2^\circ$ & $1.2\%$\\
Mirror Transmission & $3\%$ & $1.5\%$ \\
Power & $4\%$ & $ 2\%$\\
Other &  & $\approx 5\%$\\
\hline  Total Uncertainty & - & $\textbf{6} \%$
\\\hline \hline
\end{tabular}
\end{center}
\caption{Systematic errors and the resulting uncertainty in the
value of the Stark transition polarizability. The ``Other" errors
include those from uncertainties in transit velocity, calculational
approximations, and residual magnetic field. These contributions
have been estimated indirectly from their effects on the ac-Stark
parameter (Table \ref{tab errorBudget}) and the correlation between
$\alpha$ and $\beta$ determined by modeling.
%\textbf{D: these are
%the errors beyond what comes from the map. We need to say something
%to this effect, and add them to errors in Eq.(\ref{Eq_beta_final}).
%Somehow, here and in the table for $\alpha$ these are not really
%"total" unc. Any suggestions ?}
} \label{tab errorBudget_beta}
\end{table}

\section{Conclusion}

The results of the measurement of the ac-Stark shift of the
\SSZeroToTDOne transition at 408 nm [Eq. \eqref{Eq_alpha_final}] are
in good agreement with estimates of the ac-Stark shift of the
\SSZero ground state [Eq. \eqref{Eq_alpha_gs_estimate}]. The value
of the ac-Stark shifts of this transition should not limit our
ability to study parity nonconserving effects: the broadening due to
ac-Stark shifts will not preclude resolving Zeeman components of the
transition as necessitated by the current PNC-experiment scheme, and
the present research has shown that lineshapes can be understood and
modeled (see Fig. \ref{fig scanFits}). In the future high-precision
PNC experiments it will be necessary to work with a higher-finesse
cavity. In that case, it may be possible to reduce the ac-Stark
shifts by operating the cavity in the confocal regime, and expanding
the laser beam to simultaneously excite a large number of the
degenerate transverse modes.

An additional result of the present work is a new independent
determination of the Stark transition polarizability $\beta$. The
result [Eq. \ref{Eq_beta_final}] is consistent with the earlier
result [Eq. \ref{eq beta}] of Ref.\ \cite{bowers99} and has
comparable uncertainty, but it does not rely on the knowledge of the
decay branching ratios of the \TDOne state.

\section{Acknowledgements}

The authors are grateful to David DeMille for initiating this
research, and to Damon English, Derek Kimball, Chih-Hao Li, Angom
Dilip, K. Tsigutkin, and Marcis Auzinsh for very useful discussions
and criticism. This work has been supported by NSF, and by the
Director, Office of Science, Office of Basic Energy Sciences,
Nuclear Science Division, of the U.S. Department of Energy under
contract DE-AC03-76SF00098.

\bibliography{JAtomicPNC}

\begin{thebibliography}{27}
\expandafter\ifx\csname natexlab\endcsname\relax\def\natexlab#1{#1}\fi
\expandafter\ifx\csname bibnamefont\endcsname\relax
  \def\bibnamefont#1{#1}\fi
\expandafter\ifx\csname bibfnamefont\endcsname\relax
  \def\bibfnamefont#1{#1}\fi
\expandafter\ifx\csname citenamefont\endcsname\relax
  \def\citenamefont#1{#1}\fi
\expandafter\ifx\csname url\endcsname\relax
  \def\url#1{\texttt{#1}}\fi
\expandafter\ifx\csname urlprefix\endcsname\relax\def\urlprefix{URL }\fi
\providecommand{\bibinfo}[2]{#2}
\providecommand{\eprint}[2][]{\url{#2}}

\bibitem[{\citenamefont{Wieman et~al.}(1987)\citenamefont{Wieman, Noecker,
  Masterson, and Cooper}}]{wieman87}
\bibinfo{author}{\bibfnamefont{C.~E.} \bibnamefont{Wieman}},
  \bibinfo{author}{\bibfnamefont{M.~C.} \bibnamefont{Noecker}},
  \bibinfo{author}{\bibfnamefont{B.~P.} \bibnamefont{Masterson}},
  \bibnamefont{and} \bibinfo{author}{\bibfnamefont{J.}~\bibnamefont{Cooper}},
  \bibinfo{journal}{Physical Review Letters} \textbf{\bibinfo{volume}{58}},
  \bibinfo{pages}{1738} (\bibinfo{year}{1987}).

\bibitem[{\citenamefont{Wood et~al.}(1999)\citenamefont{Wood, Bennett, Roberts,
  Cho, and Wieman}}]{wood99}
\bibinfo{author}{\bibfnamefont{C.~S.} \bibnamefont{Wood}},
  \bibinfo{author}{\bibfnamefont{S.~C.} \bibnamefont{Bennett}},
  \bibinfo{author}{\bibfnamefont{J.~L.} \bibnamefont{Roberts}},
  \bibinfo{author}{\bibfnamefont{D.}~\bibnamefont{Cho}}, \bibnamefont{and}
  \bibinfo{author}{\bibfnamefont{C.~E.} \bibnamefont{Wieman}},
  \bibinfo{journal}{Canadian Journal of Physics} \textbf{\bibinfo{volume}{77}},
  \bibinfo{pages}{7} (\bibinfo{year}{1999}).

\bibitem[{\citenamefont{DeMille}(1995)}]{demille95}
\bibinfo{author}{\bibfnamefont{D.}~\bibnamefont{DeMille}},
  \bibinfo{journal}{Physical Review Letters} \textbf{\bibinfo{volume}{74}},
  \bibinfo{pages}{4165} (\bibinfo{year}{1995}).

\bibitem[{\citenamefont{Bowers et~al.}(1999)\citenamefont{Bowers, Budker,
  Freedman, Gwinner, Stalnaker, and DeMille}}]{bowers99}
\bibinfo{author}{\bibfnamefont{C.~J.} \bibnamefont{Bowers}},
  \bibinfo{author}{\bibfnamefont{D.}~\bibnamefont{Budker}},
  \bibinfo{author}{\bibfnamefont{S.~J.} \bibnamefont{Freedman}},
  \bibinfo{author}{\bibfnamefont{G.}~\bibnamefont{Gwinner}},
  \bibinfo{author}{\bibfnamefont{J.~E.} \bibnamefont{Stalnaker}},
  \bibnamefont{and} \bibinfo{author}{\bibfnamefont{D.}~\bibnamefont{DeMille}},
  \bibinfo{journal}{Physical Review A} \textbf{\bibinfo{volume}{59}},
  \bibinfo{pages}{3513} (\bibinfo{year}{1999}).

\bibitem[{\citenamefont{Stalnaker et~al.}(2002)\citenamefont{Stalnaker, Budker,
  DeMille, Freedman, and Yashchuk}}]{stalnaker02}
\bibinfo{author}{\bibfnamefont{J.~E.} \bibnamefont{Stalnaker}},
  \bibinfo{author}{\bibfnamefont{D.}~\bibnamefont{Budker}},
  \bibinfo{author}{\bibfnamefont{D.~P.} \bibnamefont{DeMille}},
  \bibinfo{author}{\bibfnamefont{S.~J.} \bibnamefont{Freedman}},
  \bibnamefont{and} \bibinfo{author}{\bibfnamefont{V.~V.}
  \bibnamefont{Yashchuk}}, \bibinfo{journal}{Physical Review A}
  \textbf{\bibinfo{volume}{66}}, \bibinfo{pages}{31403} (\bibinfo{year}{2002}).

\bibitem[{\citenamefont{Budker and Stalnaker}(2003)}]{budker03}
\bibinfo{author}{\bibfnamefont{D.}~\bibnamefont{Budker}} \bibnamefont{and}
  \bibinfo{author}{\bibfnamefont{J.~E.} \bibnamefont{Stalnaker}},
  \bibinfo{journal}{Physical Review Letters} \textbf{\bibinfo{volume}{91}},
  \bibinfo{pages}{263901/1} (\bibinfo{year}{2003}).

\bibitem[{\citenamefont{Happer}(1970)}]{happer70}
\bibinfo{author}{\bibfnamefont{W.}~\bibnamefont{Happer}},
  \bibinfo{journal}{Progress in Quantum Electronics}
  \textbf{\bibinfo{volume}{1}}, \bibinfo{pages}{47} (\bibinfo{year}{1970}).

\bibitem[{\citenamefont{Dupont-Roc and J.}(1972)}]{cohentannoudji72}
\bibinfo{author}{\bibfnamefont{C.~C.-T.} \bibnamefont{Dupont-Roc}}
  \bibnamefont{and} \bibinfo{author}{\bibnamefont{J.}},
  \bibinfo{journal}{Physical Review A} \textbf{\bibinfo{volume}{5}},
  \bibinfo{pages}{968} (\bibinfo{year}{1972}).

\bibitem[{\citenamefont{Cho et~al.}(1997)\citenamefont{Cho, Wood, Bennett,
  Roberts, and Wieman}}]{cho97}
\bibinfo{author}{\bibfnamefont{D.}~\bibnamefont{Cho}},
  \bibinfo{author}{\bibfnamefont{C.~S.} \bibnamefont{Wood}},
  \bibinfo{author}{\bibfnamefont{S.~C.} \bibnamefont{Bennett}},
  \bibinfo{author}{\bibfnamefont{J.~L.} \bibnamefont{Roberts}},
  \bibnamefont{and} \bibinfo{author}{\bibfnamefont{C.~E.}
  \bibnamefont{Wieman}}, \bibinfo{journal}{Physical Review A}
  \textbf{\bibinfo{volume}{55}}, \bibinfo{pages}{1007} (\bibinfo{year}{1997}).

\bibitem[{\citenamefont{Park et~al.}(2001)\citenamefont{Park, Noh, Lee, and
  Cho}}]{park01}
\bibinfo{author}{\bibfnamefont{C.~Y.} \bibnamefont{Park}},
  \bibinfo{author}{\bibfnamefont{H.}~\bibnamefont{Noh}},
  \bibinfo{author}{\bibfnamefont{C.~M.} \bibnamefont{Lee}}, \bibnamefont{and}
  \bibinfo{author}{\bibfnamefont{D.}~\bibnamefont{Cho}},
  \bibinfo{journal}{Physical Review A} \textbf{\bibinfo{volume}{63}},
  \bibinfo{pages}{032512/1} (\bibinfo{year}{2001}).

\bibitem[{\citenamefont{Park et~al.}(2002)\citenamefont{Park, Ji~Young,
  Jong~Min, and Cho}}]{park02}
\bibinfo{author}{\bibfnamefont{C.~Y.} \bibnamefont{Park}},
  \bibinfo{author}{\bibfnamefont{K.}~\bibnamefont{Ji~Young}},
  \bibinfo{author}{\bibfnamefont{S.}~\bibnamefont{Jong~Min}}, \bibnamefont{and}
  \bibinfo{author}{\bibfnamefont{D.}~\bibnamefont{Cho}},
  \bibinfo{journal}{Physical Review A} \textbf{\bibinfo{volume}{65}},
  \bibinfo{pages}{033410/1} (\bibinfo{year}{2002}).

\bibitem[{\citenamefont{Sobelman}(1992)}]{sobelman92}
\bibinfo{author}{\bibfnamefont{I.~I.} \bibnamefont{Sobelman}},
  \emph{\bibinfo{title}{Atomic Spectra and Radiative Transitions}}, Springer
  Series on Atoms and Plasmas (\bibinfo{publisher}{Springer},
  \bibinfo{address}{New York}, \bibinfo{year}{1992}), \bibinfo{edition}{2nd}
  ed.

\bibitem[{\citenamefont{Varshalovich et~al.}(1988)\citenamefont{Varshalovich,
  Moskalev, and Khersonskii}}]{Var88}
\bibinfo{author}{\bibfnamefont{D.~A.} \bibnamefont{Varshalovich}},
  \bibinfo{author}{\bibfnamefont{A.~N.} \bibnamefont{Moskalev}},
  \bibnamefont{and} \bibinfo{author}{\bibfnamefont{V.~K.}
  \bibnamefont{Khersonskii}}, \emph{\bibinfo{title}{Quantum theory of angular
  momentum: irreducible tensors, spherical harmonics, vector coupling
  coefficients, 3nj symbols}} (\bibinfo{publisher}{World Scientific},
  \bibinfo{address}{Singapore}, \bibinfo{year}{1988}).

\bibitem[{\citenamefont{Porsev et~al.}(1999)\citenamefont{Porsev, Rakhlina~Yu,
  and Kozlov}}]{porsev99}
\bibinfo{author}{\bibfnamefont{S.~G.} \bibnamefont{Porsev}},
  \bibinfo{author}{\bibfnamefont{G.}~\bibnamefont{Rakhlina~Yu}},
  \bibnamefont{and} \bibinfo{author}{\bibfnamefont{M.~G.}
  \bibnamefont{Kozlov}}, \bibinfo{journal}{Physical Review A}
  \textbf{\bibinfo{volume}{60}}, \bibinfo{pages}{2781} (\bibinfo{year}{1999}).

\bibitem[{\citenamefont{Blagoev and Komarovskii}(1994)}]{blagoev94}
\bibinfo{author}{\bibfnamefont{K.~B.} \bibnamefont{Blagoev}} \bibnamefont{and}
  \bibinfo{author}{\bibfnamefont{V.~A.} \bibnamefont{Komarovskii}},
  \bibinfo{journal}{Atomic Data and Nuclear Data Tables}
  \textbf{\bibinfo{volume}{56}}, \bibinfo{pages}{1} (\bibinfo{year}{1994}).

\bibitem[{\citenamefont{Martin et~al.}(1978)\citenamefont{Martin, Zalubas, and
  Hagan}}]{martin78}
\bibinfo{author}{\bibfnamefont{W.~C.} \bibnamefont{Martin}},
  \bibinfo{author}{\bibfnamefont{R.}~\bibnamefont{Zalubas}}, \bibnamefont{and}
  \bibinfo{author}{\bibfnamefont{L.}~\bibnamefont{Hagan}},
  \bibinfo{journal}{Nat. Bur. Standards. 1978} pp. \bibinfo{pages}{vii+411}
  (\bibinfo{year}{1978}).

\bibitem[{\citenamefont{Anderson et~al.}(1984)\citenamefont{Anderson, Frisch,
  and Masser}}]{anderson84}
\bibinfo{author}{\bibfnamefont{D.~Z.} \bibnamefont{Anderson}},
  \bibinfo{author}{\bibfnamefont{J.~C.} \bibnamefont{Frisch}},
  \bibnamefont{and} \bibinfo{author}{\bibfnamefont{C.~S.}
  \bibnamefont{Masser}}, \bibinfo{journal}{Applied Optics}
  \textbf{\bibinfo{volume}{23}}, \bibinfo{pages}{1238} (\bibinfo{year}{1984}).

\bibitem[{\citenamefont{Hood et~al.}(2001)\citenamefont{Hood, Kimble, and
  Ye}}]{hood01}
\bibinfo{author}{\bibfnamefont{C.~J.} \bibnamefont{Hood}},
  \bibinfo{author}{\bibfnamefont{H.~J.} \bibnamefont{Kimble}},
  \bibnamefont{and} \bibinfo{author}{\bibfnamefont{J.}~\bibnamefont{Ye}},
  \bibinfo{journal}{Physical Review A} \textbf{\bibinfo{volume}{64}},
  \bibinfo{pages}{033804/1} (\bibinfo{year}{2001}).

\bibitem[{\citenamefont{Drever et~al.}(1983)\citenamefont{Drever, Hall,
  Kowalski, Hough, Ford, Munley, and Ward}}]{drever83}
\bibinfo{author}{\bibfnamefont{R.~W.~P.} \bibnamefont{Drever}},
  \bibinfo{author}{\bibfnamefont{J.~L.} \bibnamefont{Hall}},
  \bibinfo{author}{\bibfnamefont{F.~V.} \bibnamefont{Kowalski}},
  \bibinfo{author}{\bibfnamefont{J.}~\bibnamefont{Hough}},
  \bibinfo{author}{\bibfnamefont{G.~M.} \bibnamefont{Ford}},
  \bibinfo{author}{\bibfnamefont{A.~J.} \bibnamefont{Munley}},
  \bibnamefont{and} \bibinfo{author}{\bibfnamefont{H.}~\bibnamefont{Ward}},
  \bibinfo{journal}{Applied Physics B-Photophysics and Laser Chemistry}
  \textbf{\bibinfo{volume}{B31}}, \bibinfo{pages}{97} (\bibinfo{year}{1983}).

\bibitem[{\citenamefont{Budker et~al.}(2000)\citenamefont{Budker, Rochester,
  and Yashchuk}}]{budker00}
\bibinfo{author}{\bibfnamefont{D.}~\bibnamefont{Budker}},
  \bibinfo{author}{\bibfnamefont{S.~M.} \bibnamefont{Rochester}},
  \bibnamefont{and} \bibinfo{author}{\bibfnamefont{V.~V.}
  \bibnamefont{Yashchuk}}, \bibinfo{journal}{Review of Scientific Instruments}
  \textbf{\bibinfo{volume}{71}}, \bibinfo{pages}{2984} (\bibinfo{year}{2000}).

\bibitem[{\citenamefont{Siegman}(1986)}]{siegman86}
\bibinfo{author}{\bibfnamefont{A.~E.} \bibnamefont{Siegman}},
  \emph{\bibinfo{title}{Lasers}} (\bibinfo{publisher}{University Science
  Books}, \bibinfo{address}{Sausalito}, \bibinfo{year}{1986}).

\bibitem[{\citenamefont{Corney}(1988)}]{Cor88}
\bibinfo{author}{\bibfnamefont{A.}~\bibnamefont{Corney}},
  \emph{\bibinfo{title}{Atomic and Laser Spectroscopy}}
  (\bibinfo{publisher}{Clarendon}, \bibinfo{address}{Oxford},
  \bibinfo{year}{1988}).

\bibitem[{\citenamefont{Press et~al.}(1999)\citenamefont{Press, Teukolsky,
  Vetterling, and Flannery}}]{press99}
\bibinfo{author}{\bibfnamefont{W.~H.} \bibnamefont{Press}},
  \bibinfo{author}{\bibfnamefont{S.~A.} \bibnamefont{Teukolsky}},
  \bibinfo{author}{\bibfnamefont{W.~T.} \bibnamefont{Vetterling}},
  \bibnamefont{and} \bibinfo{author}{\bibfnamefont{B.~P.}
  \bibnamefont{Flannery}}, \emph{\bibinfo{title}{Numerical Recipies in C: The
  Art of Scientific Computing}} (\bibinfo{publisher}{Cambridge University
  Press}, \bibinfo{address}{New York}, \bibinfo{year}{1999}),
  \bibinfo{edition}{2nd} ed.

\bibitem[{\citenamefont{Ramsey}(1990)}]{ramsey56}
\bibinfo{author}{\bibfnamefont{N.~F.} \bibnamefont{Ramsey}},
  \emph{\bibinfo{title}{Molecular Beams}} (\bibinfo{publisher}{Oxford
  University Press}, \bibinfo{address}{New York}, \bibinfo{year}{1990}).

\bibitem[{\citenamefont{Arfken and Weber}(2005)}]{Arf2005}
\bibinfo{author}{\bibfnamefont{G.~B.} \bibnamefont{Arfken}} \bibnamefont{and}
  \bibinfo{author}{\bibfnamefont{H.-J.} \bibnamefont{Weber}},
  \emph{\bibinfo{title}{Mathematical methods for physicists}}
  (\bibinfo{publisher}{Elsevier}, \bibinfo{address}{Boston},
  \bibinfo{year}{2005}), \bibinfo{edition}{6th} ed.

\bibitem[{\citenamefont{Regan}(2001)}]{Reg2001}
\bibinfo{author}{\bibfnamefont{B.~C.} \bibnamefont{Regan}},
  \bibinfo{type}{Ph.d. thesis}, \bibinfo{school}{University of California}
  (\bibinfo{year}{2001}), \bibinfo{note}{available from the author or
  http://www.umi.com}.

\bibitem[{\citenamefont{Wood et~al.}(1997)\citenamefont{Wood, Bennett, Cho,
  Masterson, Roberts, Tanner, and Wieman}}]{wood97}
\bibinfo{author}{\bibfnamefont{C.~S.} \bibnamefont{Wood}},
  \bibinfo{author}{\bibfnamefont{S.~C.} \bibnamefont{Bennett}},
  \bibinfo{author}{\bibfnamefont{D.}~\bibnamefont{Cho}},
  \bibinfo{author}{\bibfnamefont{B.~P.} \bibnamefont{Masterson}},
  \bibinfo{author}{\bibfnamefont{J.~L.} \bibnamefont{Roberts}},
  \bibinfo{author}{\bibfnamefont{C.~E.} \bibnamefont{Tanner}},
  \bibnamefont{and} \bibinfo{author}{\bibfnamefont{C.~E.}
  \bibnamefont{Wieman}}, \bibinfo{journal}{Science}
  \textbf{\bibinfo{volume}{275}}, \bibinfo{pages}{1759} (\bibinfo{year}{1997}).

\end{thebibliography}

\end{document}